\documentclass[11pt]{article}

\usepackage{graphicx}
\usepackage{amsmath}
\usepackage{dcolumn}
\usepackage{bm}
\usepackage{slashed}
\usepackage[bookmarks,bookmarksnumbered,colorlinks=true,anchorcolor=blue,
linkcolor=blue,urlcolor=blue,citecolor=blue,breaklinks=true]{hyperref}
\usepackage[utf8]{inputenc}
\usepackage{authblk}
\usepackage{cite}
\usepackage{titlesec}
\usepackage{caption}
\usepackage{subcaption}
\usepackage{float}
\usepackage{color}
\usepackage{ulem}

\newcommand{\be}{\begin{equation}}
\newcommand{\ee}{\end{equation}}
\newcommand{\ba}{\begin{eqnarray}}
\newcommand{\ea}{\end{eqnarray}}
\def\bea{\begin{eqnarray}}
\def\eea{\end{eqnarray}}

\newcommand{\gsim}{\mathrel{\hbox{\rlap{\lower.55ex \hbox {$\sim$}}
                   \kern-.3em \raise.4ex \hbox{$>$}}}}
\newcommand{\lsim}{\mathrel{\hbox{\rlap{\lower.55ex \hbox {$\sim$}}
                   \kern-.3em \raise.4ex \hbox{$<$}}}}

\def\roughly#1{\mathrel{\raise.3ex\hbox{$#1$\kern-.75em%
\lower1ex\hbox{$\sim$}}}}
\def\lsim{\roughly<}
\def\gsim{\roughly>}

\def\({\left(}
\def\){\right)}
\def\[{\left[}
\def\]{\right]}
\def\<{\langle}
\def\>{\rangle}

\usepackage{xcolor}

\begin{document}
\title{\bf The holographic entanglement pattern of BTZ planar black hole from a thread perspective}

\author[]{Yi-Yu Lin$^{1,2}$ \thanks{yiyu@simis.cn}}
\author[]{Dong-Yu Fang$^{3}$ \thanks{fangdongyu42@gmail.com}}
\author[]{Jie-Chen Jin$^{4}$ \thanks{Jinjch5@mail3.sysu.edu.cn}}
\author[]{Chen-Ye Li$^{3}$ \thanks{endgamelichenye@gmail.com}}

 \affil{${}^1$Fudan Center for Mathematics and Interdisciplinary Study, Fudan University, Shanghai, 200433, China}
\affil[]{${}^2$Shanghai Institute for Mathematics and Interdisciplinary Sciences (SIMIS), Shanghai, 200433, China}
 \affil{${}^3$Department of Physics , Fudan University, Shanghai, 200433, China}
  \affil{${}^4$Department of Physics and Astronomy, University College London,  Gower Street, London, WC1E 6BT, UK}


\maketitle

\begin{abstract}

In this paper, we study the holographic quantum entanglement structure in the finite-temperature CFT state/planar BTZ black hole correspondence from the perspective of entanglement threads. Unlike previous studies based on bit threads, these entanglement threads provide a more detailed characterization of the contribution sources to the von Neumann entropy of boundary subregions, in particular by quantitatively deriving the flux function of entanglement threads that traverse the wormhole horizon and connect the two asymptotic boundaries. Since entanglement threads are naturally and closely related to tensor network states, the results are argued to imply the existence of the perfect-type entanglement formed jointly by the entanglement threads crossing the wormhole and the internal threads in the single-sided boundary. We also discuss the close connections of this work with concepts such as bit threads and partial entanglement entropy.

\end{abstract}
\tableofcontents

\newpage

\section{Introduction}

\begin{figure}\label{0.1}
     \centering
     \begin{subfigure}[b]{0.24\textwidth}
         \centering
         \includegraphics[width=\textwidth]{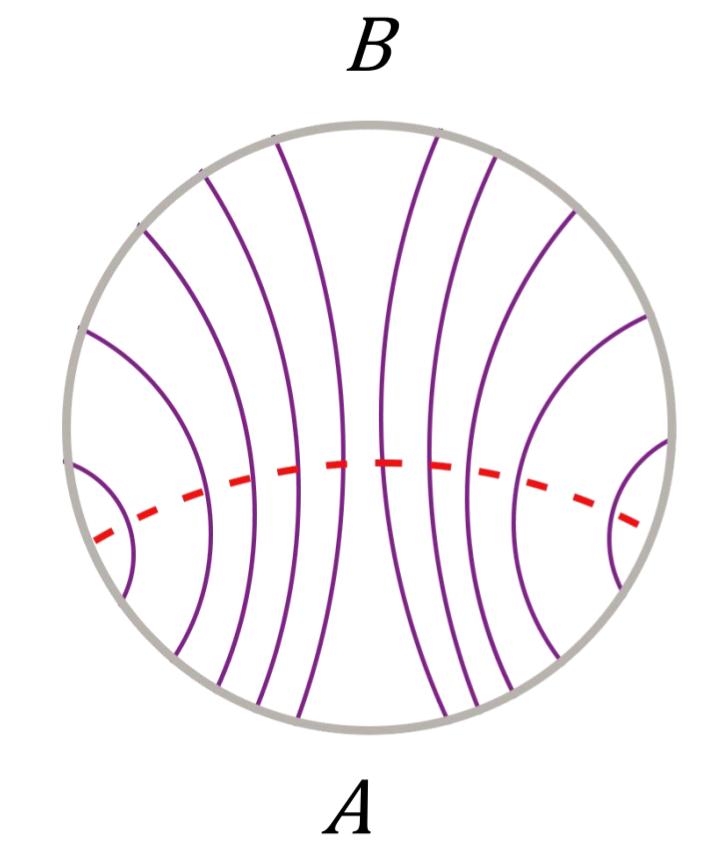}
         \caption{}
         \label{0.1a}
     \end{subfigure}
     \hfill
     \begin{subfigure}[b]{0.5\textwidth}
         \centering
         \includegraphics[width=\textwidth]{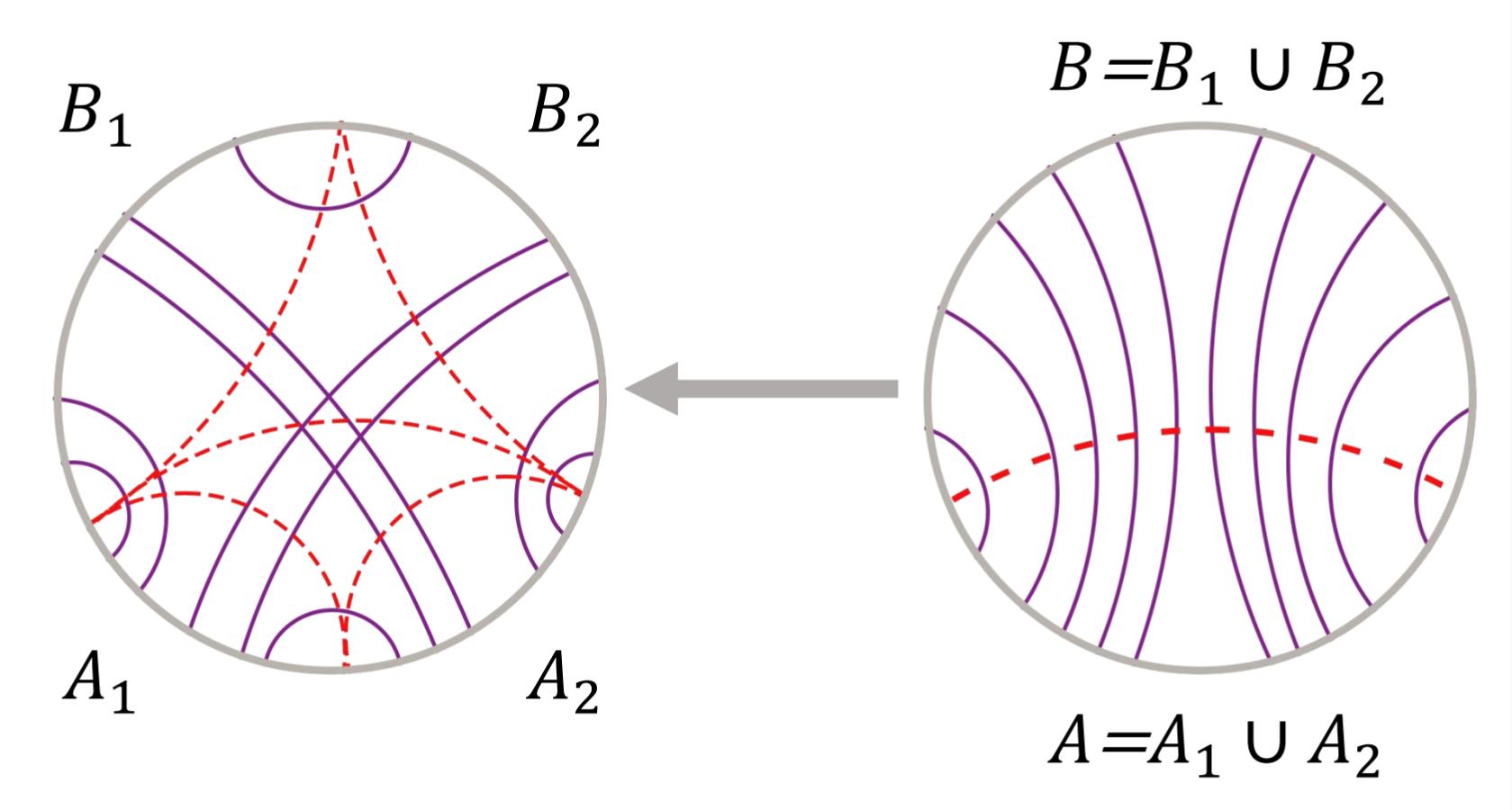}
         \caption{}
         \label{0.1b}
     \end{subfigure}
        \caption{(a) A holographic ``thread'' picture characterizing the entanglement entropy between two complementary regions. (b) A more refined thread configuration characterizing a set of entanglement entropies involving more subregions. By iteratively dividing the quantum system, more and more refined thread configurations can be constructed, characterizing the entanglement structure at more and more refined levels.  Here the RT surfaces are represented as red dashed lines, while the threads are schematically represented as purple lines.}
\end{figure}

To clarify what we aim to do and how it might offer insights into the holographic entanglement structure of black hole (or wormhole) spacetimes, we begin with a background introduction using the pure AdS/vacuum CFT duality as an example. For a long time, tensor network tools~\cite{Swingle:2009bg, Swingle:2012wq, Pastawski:2015qua, Vidal:2007hda, Vidal:2008zz, Vidal:2015} have been predominantly used to study the mechanism of entanglement structure in holographic duality~\cite{Maldacena:1997re,Gubser:1998bc,Witten:1998qj}. However, there are also studies that approach this topic from an explicit ``thread picture'' perspective (see~\cite{Freedman:2016zud, Cui:2018dyq, Headrick:2017ucz, Headrick:2022nbe, Agon:2018lwq, Kudler-Flam:2019oru, Lin:2025yko, Lin:2023jah,Lin:2022flo,Lin:2022agc,Lin:2020yzf,Lin:2023orb,Lin:2021hqs, Jahn:2019nmz, Yan:2019vzp, Lin:2024dho,Lin:2023rxc, Caggioli:2024uza, Mintchev:2022fcp, Lin:2023hzs, Lin:2022aqf, Wen:2024uwr,Headrick:2020gyq} for examples). Although these perspectives differ slightly, they essentially express a rather natural (albeit kind of naive) idea of refining the decomposition of holographic entanglement entropy. This idea can be intuitively described by the following narrative: we imagine representing quantum entanglement in a holographic system as a picture of threads. Then, for the Ryu-Takayanagi formula~\cite{Ryu:2006bv,Ryu:2006ef,Hubeny:2007xt}, we obtain a naïve diagram as shown in Fig.~\ref{0.1a}. The earliest origin of this type of holographic diagram can be traced back to bit threads~\cite{Freedman:2016zud,Cui:2018dyq,Headrick:2017ucz,Headrick:2022nbe}. More concretely, consider the entanglement entropy between a subregion $A$ and its complement $B$ in a pure state of a CFT dual to pure AdS. We may envision a family of continuous bulk curves (i.e., threads), each connecting $A$ and its complement $B$, passing through the RT surface $\gamma_A$, with the number of threads exactly equal to the entanglement entropy between $A$ and $B$:
\begin{equation}\label{fab}
F_{AB} = S_A
\end{equation}
We appropriately refer to these threads as \emph{holographic entanglement threads}, or simply \emph{entanglement threads}~\cite{ Lin:2025yko}.

One can do better by constructing a series of increasingly refined thread configurations that can be used to compute the entanglement entropies of more than one region. As shown in Fig.~\ref{0.1b}, we further decompose region $A$ into $A = A_1 \cup A_2$, and region $B$ into $B = A_3 \cup A_4$. Then, we can construct a more refined thread configuration capable of computing the entanglement entropies between six connected subregions and their complements, including: $A_1$, $A_2$, $A_3$, $A_4$, $A = A_1 \cup A_2$, and $A_2 \cup B_3$. In other words, the number of threads connecting each of these six regions to their complements is exactly equal to their corresponding entanglement entropies. 
Similarly, one can further partition the quantum system $M$ into more neighboring, non-overlapping elementary regions $A_1$, $A_2$, $\ldots$, $A_N$, and then obtain a correspondingly finer thread configuration. This process can be iterated (as long as each elementary region remains much larger than the Planck scale, ensuring the applicability of the RT formula). Here, we define the elementary regions to satisfy: $A_i \cap A_j = \emptyset,\quad \bigcup A_i = M.$ Next, for each pair of elementary regions $A_i$ and $A_j$, we define a function $F_{ij} \equiv F_{A_i \leftrightarrow A_j}$, which denotes the number of threads connecting $A_i$ and $A_j$. We then impose the following physical requirement on the thread configuration $\{F_{ij}\}$~\cite{Lin:2021hqs}:
\begin{equation}
S_{a(a+1)\ldots b} = \sum_{i,j} F_{ij}, \quad \text{where } i \in \{a,a+1,\ldots,b\},\; j \notin \{a,a+1,\ldots,b\} \label{equ}.
\end{equation}
Here, $S_{a(a+1)\ldots b}$ denotes the entanglement entropy $S_A$ of a connected composite region $A = A_{a(a+1)\cdots b} \equiv A_a \cup A_{a+1} \cup \cdots \cup A_b$. This equation can be intuitively understood: the entanglement entropy between $A$ and $\bar A$ arises from the sum of all $F_{ij}$, where $A_i$ is within $A$ and $A_j$ is within its complement $\bar A$. Solving equation~\eqref{equ}, the first thing we discover is that the number of threads connecting two elementary regions is exactly given by half the so-called conditional mutual information~\cite{Lin:2021hqs}. In other words, conditional mutual information characterizes the correlation between two regions $A_i$ and $A_j$ separated by a distance $L$:
\begin{equation}
\begin{aligned}
F_{A_i \leftrightarrow A_j} &= \frac{1}{2} I(A_i, A_j | \tilde{L}) \\
&\equiv \frac{1}{2} \big[S(A_i \cup \tilde{L}) + S(A_j \cup \tilde{L}) - S(A_i \cup \tilde{L} \cup A_j) - S(\tilde{L})\big]
\end{aligned} \label{cmi}
\end{equation}
Here, we denote the region between $A_i$ and $A_j$ as $\tilde{L} = A_{(i+1)\cdots(j-1)}$, which is a composite region consisting of many elementary regions and represents the distance between $A_i$ and $A_j$.

In fact, this kind of refined thread configurations is closely related to various concepts proposed from different perspectives in the study of holographic duality, including kinematic space~\cite{Czech:2015kbp,Czech:2015qta}, entropy cone~\cite{Bao:2015bfa,Hubeny:2018ijt,Hubeny:2018trv,HernandezCuenca:2019wgh}, and holographic partial entanglement entropy~\cite{Vidal:2014aal,Wen:2019iyq,Wen:2018whg,Kudler-Flam:2019oru}. These connections are, in some sense, natural and easily obtained—especially considering that conditional mutual information plays a central role in all of them. In kinematic space~\cite{Czech:2015kbp,Czech:2015qta}, it is defined as the volume measure, thus characterizing the density of entanglement entropy. Discussions on these connections can be found in a series of papers~\cite{Lin:2025yko, Lin:2022flo, Lin:2022agc, Lin:2023orb, Lin:2023jah}. In particular, in~\cite{Lin:2025yko}, two key points have been emphasized. First, at least in the case of pure AdS/vacuum CFT, the trajectories of these entanglement threads in the holographic bulk are exactly geodesics (see also~\cite{Agon:2018lwq, Lin:2024dho}). Therefore, each entanglement thread has a one-to-one correspondence with a point in the effective kinematic space.\footnote{The term ``effective'' is used here because the full kinematic space contains redundant regions that represent the same undirected geodesic information.} Second, these entanglement threads can be used to describe tensor network states. Since tensor network states can be expressed as quantum circuits\cite{Pastawski:2015qua}, ~\cite{Lin:2025yko} demonstrate that these entanglement threads exactly form the collection of wires in such a quantum circuit. This latter point allows us to meaningfully assign the concept of quantum states to these entanglement threads. As we will see later, the connection between entanglement threads and quantum states can be used to characterize phase transition behaviors of holographic RT surface calculating entanglement entropy, thereby capturing the features of the system's entanglement structure.

In this paper, we aim to generalize the previously well-studied entanglement thread picture in pure AdS/CFT vacuum~\cite{Lin:2025yko, Lin:2022flo, Lin:2023orb}
 to the case involving AdS black holes/finite-temperature CFT duals. This not only helps us analyze the entanglement structure of finite-temperature thermal states, especially through the notion of partial entanglement entropy~\cite{Vidal:2014aal,Wen:2019iyq,Wen:2018whg,Kudler-Flam:2019oru}, but also facilitates our understanding of the entanglement structure of black hole spacetime geometries. Moreover, this study serves as a paradigmatic generalization of the entanglement thread picture from pure states to mixed states. This is especially significant considering that, in mixed state systems, the von Neumann entropy of a subregion typically does not equal that of its spatial complement, which presents a conceptual challenge to the idea of entanglement threads.

The structure of this paper is as follows: In Section~\ref{sec2}, we review certain concepts and results related to holographic conditional mutual information in the pure AdS case and propose a preliminary extension to the planar BTZ black hole case. In Section~\ref{sec3}, we further propose a concrete scheme for the entanglement thread structure in the planar BTZ black hole case. This is a more refined proposal that utilizes the exact correspondence between the two-sided planar BTZ black hole slice and the Poincaré disk. In Section~\ref{sec4}, we discuss the close connections and improvements of our results compared with previous literature. Section~\ref{sec5} concludes the paper.

\section{Background Review and Preliminary Proposal}\label{sec2}

\subsection{Conditional Mutual Information in Pure AdS}\label{sec21}

As mentioned in the introduction, within the thread picture of AdS/CFT duality, the quantity known as conditional mutual information plays a central role in quantum information theory. For the pure AdS/vacuum CFT correspondence, it has been thoroughly studied at various levels, and its expression is well known. We recall its expression here (which is not new, as it appears in various references, e.g.,~\cite{Czech:2015qta,Kudler-Flam:2019oru,Wen:2018whg, Asplund:2016koz, Levay:2020rzs, Boldis:2021snw}), but we offer a novel and interesting comment from the perspective of the thread picture: since conditional mutual information, as we shall see, is a finite quantity independent of the cutoff, it can in fact serve as a starting point to provide a regularization scheme for holographic entanglement entropy.

The conditional mutual information $I(A_i, A_j|\tilde{L})$, half of which we identify as the ``number of entanglement threads" $F_{ij}$, has already been defined as~(\ref{cmi}) in the introduction. Here we recall the expression of $I(A_i, A_j|\tilde{L})$ or equivalently $F_{ij}$ on a constant time slice of pure $\mathrm{AdS}_3$ (i.e.,  a Poincaré disk). First, we review some basic facts about $\mathrm{AdS}_3$. Begin with the embedding trajectory of $\mathrm{AdS}_3$ spacetime in higher dimensions:
\begin{equation}
-({X^0})^2 + ({X^1})^2 + ({X^2})^2 - ({X^3})^2 = -L^2 \label{tra},
\end{equation}
where $L$ is the AdS radius. Parametrize it as:
\begin{equation}
\begin{aligned}
X^0 &= L\cosh \theta \cos \varphi, \\
X^3 &= L\cosh \theta \sin \varphi, \\
X^1 &= \sin \xi \cdot L\sinh \theta, \\
X^2 &= \cos \xi \cdot L\sinh \theta.
\end{aligned}
\end{equation}
This leads to the typical form of the global coordinates in $\mathrm{AdS}_3$ spacetime:
\begin{equation}
ds^2 = L^2(-\cosh^2\theta\, d\varphi^2 + d\theta^2 + \sinh^2\theta\, d\xi^2) \label{glo}
\end{equation}
Here, $\varphi$ represents the time coordinate, $\theta \in [0, +\infty)$ characterizes the radial coordinate (which corresponds to the radial direction of the Poincaré disk via equation~(\ref{rho})), and $\xi \in [0, 2\pi]$ characterizes the spatial angular coordinate (corresponding to the angular direction of the Poincaré disk via equation~(\ref{thet})). Taking an equal-time slice by setting $\varphi = 0$ (i.e., $X^3 = 0$), we obtain a two-dimensional Poincaré disk of radius $R$, which can be represented in Cartesian coordinates $(x, y)$ or polar coordinates $(\rho, \vartheta)$, where
\begin{equation}
x = \frac{R X^2}{L + X^0}, \quad y = \frac{R X^1}{L + X^0},
\end{equation}
so that
\begin{equation}
\rho = \sqrt{x^2 + y^2} = R\tanh\left(\frac{\theta}{2}\right) \label{rho},
\end{equation}
and
\begin{equation}
\tan \vartheta = \frac{y}{x} = \tan \xi \Rightarrow \vartheta = \xi \label{thet},
\end{equation}
 According to this, the metric becomes conformally flat:
\begin{equation}
ds^2 = \frac{4L^2 R^2}{(R^2 - \rho^2)^2}(d\rho^2 + \rho^2 d\vartheta^2) \label{conf}
,\end{equation}
where $x^2 + y^2 < R^2$, meaning that $\rho \to R$ (i.e., $\theta \to +\infty$) corresponds to the boundary of the disk.

The length of Ryu–Takayanagi geodesics in global coordinates $(\theta, \xi)$ is known in the literature~\cite{Ryu:2006bv}. The geodesic length between two points with angular coordinates $\xi_1$ and $\xi_2$ at a fixed large radial cutoff $\theta = \theta_0$ is
\begin{equation}
l(\theta_0, \xi_1;\, \theta_0, \xi_2) = 2L \cdot \ln \left(e^{\theta_0} \sin\left(\frac{\Delta \xi}{2}\right)\right) \label{lglo},
\end{equation}
where $\Delta \xi = |\xi_1 - \xi_2|$ represents the interval length of subregion $A$. Using the holographic relation~\cite{Brown:1986nw}
\begin{equation}
\frac{c}{3} = \frac{L}{2G} \label{cl},
\end{equation}
and the regularized version of the RT formula~\cite{Ryu:2006bv}
\begin{equation}
S_A^{\text{reg}} = \frac{l(\theta_0, \xi_1;\, \theta_0, \xi_2)}{4G_N} \label{reg},
\end{equation}
~(\ref{lglo}) is expected to reproduce the holographic entanglement entropy of a subregion of length $d = R\Delta\xi$ in a system of total length $\Sigma = 2\pi R$~\cite{Calabrese:2004eu}:
\begin{equation}
S_A = \frac{c}{3} \ln \left(\frac{\Sigma}{\pi\varepsilon} \sin\left(\frac{\pi d}{\Sigma}\right)\right) = \frac{c}{3} \ln \left(\frac{2R}{\varepsilon} \sin\left(\frac{\Delta\xi}{2}\right)\right) \label{en1}
\end{equation}
Of course, perfect matching requires assuming the cutoff relation~\cite{Ryu:2006bv}
\begin{equation}
e^{-\theta_0} \sim \frac{\varepsilon}{2R} \label{match}.
\end{equation}

\begin{figure}[H]
    \centering
    \includegraphics[scale=0.18]{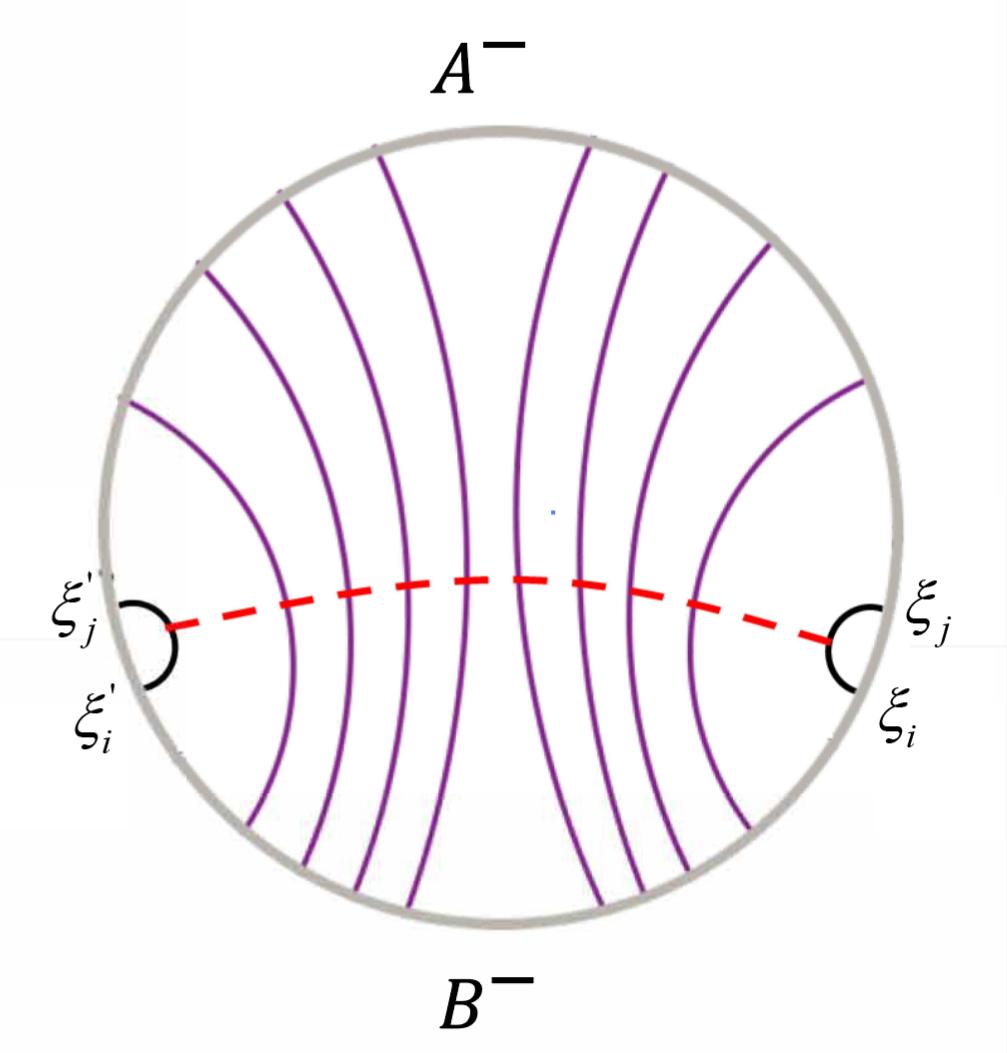}
    \caption{ By introducing two small ``gap regions", the entanglement entropy between two complementary region $A$ and $B$ can be regularized using finite CMI.}
    \label{1.2}
\end{figure}

From equation~\eqref{reg} or \eqref{en1}, we can now directly compute $I(A_i, A_j|\tilde{A})$ or $F_{ij}$ using equation~\eqref{cmi}. Let the sizes of $A_i$, $A_j$, and the intermediate region $\tilde{A}$ be $a_i = R(\xi_{i_2} - \xi_{i_1}), \quad a_j = R(\xi_{j_2} - \xi_{j_1}), \quad \tilde{a} = R(\xi_{j_1} - \xi_{i_2}),$with the ordering $\xi_{i_2} > \xi_{i_1} > \xi_{j_2} > \xi_{j_1}$. Then we obtain:
\begin{equation}
\begin{aligned}
F_{ij} &= \frac{1}{2}I(A_i, A_j|\tilde{A}) = \frac{c}{6} \ln \left(\frac{
\sin\left(\frac{\pi(a_i + \tilde{a})}{\Sigma}\right)\sin\left(\frac{\pi(a_j + \tilde{a})}{\Sigma}\right)}{
\sin\left(\frac{\pi(a_i + \tilde{a} + a_j)}{\Sigma}\right)\sin\left(\frac{\pi \tilde{a}}{\Sigma}\right)}\right) \\
&= \frac{c}{6} \ln \left(\frac{
\sin\left(\frac{|\xi_{i_1} - \xi_{j_1}|}{2}\right) \sin\left(\frac{|\xi_{i_2} - \xi_{j_2}|}{2}\right)}{
\sin\left(\frac{|\xi_{i_2} - \xi_{j_1}|}{2}\right) \sin\left(\frac{|\xi_{i_1} - \xi_{j_2}|}{2}\right)}\right)
\end{aligned} \label{fij1}.
\end{equation}
A noteworthy fact is that $F_{ij}$, or conditional mutual information, is finite; in other words, the cutoff dependence cancels out in its expression. Since $F_{ij}$ represents the number of entanglement threads in our framework, we can use this interesting fact to propose a regularization scheme for holographic entanglement entropy based on the ``number of entanglement threads”, which is particularly useful for our current study. As illustrated in FIG.~\ref{1.2}, suppose we want to regularize the entanglement entropy between a region $A$ with angular size $\Delta \xi = \xi_j - \xi_i$ and its complement $B$, which is a divergent quantity. We introduce two small ``gap regions" of diameter $\varepsilon$ (corresponding to angular width $\frac{\varepsilon}{R}$) at both ends of $A$ and $B$, and denote the remaining main parts as $A^-$ and $B^-$. The key point is that the number of entanglement threads connecting $A^-$ and $B^-$ is finite, and from equation~\eqref{fij1}, we have:
\begin{equation}
\begin{aligned}
F_{ij} &= \frac{1}{2} I(A^-, B^- | \varepsilon) = \frac{c}{6} \ln \left(\frac{
\sin\left(\frac{\Delta\xi}{2}\right)^2}{
\sin\left(\frac{\varepsilon}{2R}\right)^2}\right) \\
&= \frac{c}{3} \ln \left(\frac{
\sin\left(\frac{\Delta\xi}{2}\right)}{
\sin\left(\frac{\varepsilon}{2R}\right)}\right) \approx \frac{c}{3} \ln \left(\frac{2R}{\varepsilon} \sin\left(\frac{\Delta\xi}{2}\right)\right)
\end{aligned} \label{cal}
\end{equation}
Thus, by regularizing the number of entanglement threads, we recover the regularized entanglement entropy expression in ~\eqref{en1}. The calculation in ~\eqref{cal} can also be interpreted as the regularization of the geodesic length in~\eqref{lglo}, and the denominator factor $\sin\left(\frac{\varepsilon}{2R}\right) \approx \frac{\varepsilon}{2R}$ provides the cutoff factor $e^{-\theta_0}$, explaining the origin of~\eqref{match}. Although the above discussion might seem circular, as we first obtained the expression for conditional mutual information as a linear combination of entanglement entropies, then used it to define entanglement entropy, this depends on one's interpretation. If we regard entanglement threads as fundamental descriptors of entanglement entropy, then our approach can be understood as beginning with the finite value expressions of ``number of threads”, and then using them to characterize entanglement entropy.

\subsection{A Preliminary Proposal For The Planar BTZ Black Hole}\label{sec22}

We aim to generalize the entanglement thread picture to scenarios involving AdS black holes/finite-temperature CFT duals~\cite{Ryu:2006bv,Ryu:2006ef}. This not only helps analyze the entanglement structure of thermal states at finite temperature—particularly via the notion of partial entanglement entropy~\cite{Vidal:2014aal}—but also aids in understanding the entanglement structure of black hole spacetime geometries.

\begin{figure}[H]
    \centering
    \includegraphics[width=0.5\linewidth]{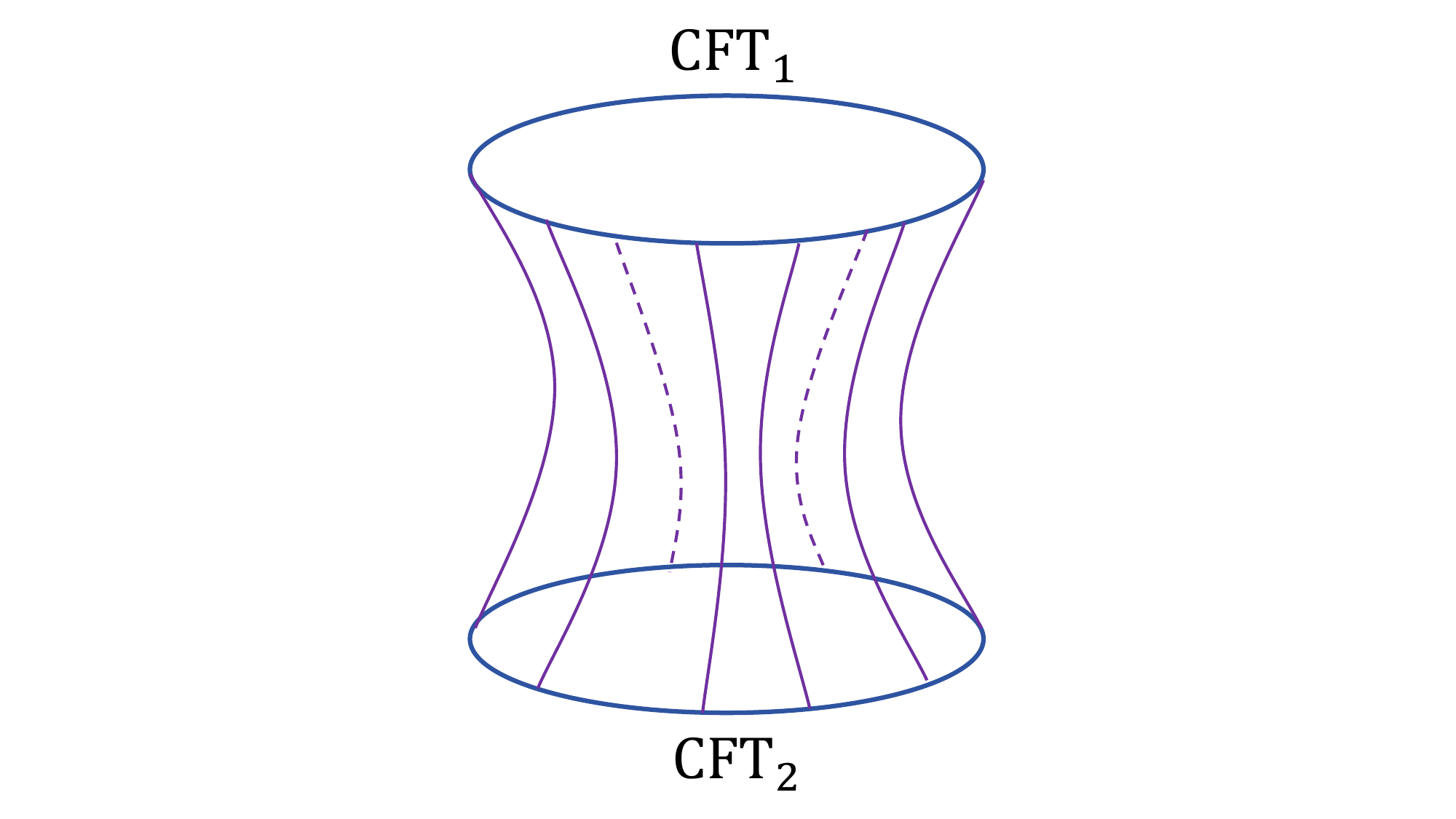}
    \caption{ By viewing the black hole as one side of a two-sided wormhole, the thread picture related to a thermal mixed state can be characterized.}
    \label{22.1}
\end{figure}

In this generalization from pure states to mixed states, we immediately encounter a conceptual difficulty: when dividing the system into two parts $A$ and $B$, the von Neumann entropy of $A$ is generally no longer equal to that of $B$. Thus, we can no longer simply apply a diagram like Fig.~\ref{0.1a} to the RT formula: it becomes unclear whether we should imagine $S(A)$ or $S(B)$ number of threads connecting $A$ and $B$. To generalize the thread picture from the vacuum state to a thermal mixed state, a natural idea is to view the black hole as one side of a two-sided wormhole~\cite{Maldacena:2001kr}. Then, the entire wormhole $W$ corresponds to a pure state—the so-called thermofield double state~\cite{Israel:1976ur} of two CFTs (denoted as CFT$_1$ and CFT$_2$). In a pure state, since the von Neumann entropy of two complementary subsystems is always equal (equal to the entanglement entropy between them), we can unambiguously assign a number of entanglement threads. In particular, as shown in Fig.~\ref{22.1}, the entanglement entropy between CFT$_1$ and CFT$_2$ can be represented simply using the thread picture. The RT surface that computes the holographic entanglement entropy $S_{\text{CFT}_1}$ is, by definition, the minimal-area extremal surface that completely separates the two CFTs, and coincides with the wormhole horizon. Our thread picture visualizes $S_{\text{CFT}_1}$ as the number of threads passing through the horizon and connecting CFT$_1$ and CFT$_2$. CFT$_1$ appears to be in a thermal state, and $S_{\text{CFT}_1}$ gives the thermal entropy of CFT$_1$ at temperature $\beta^{-1}$.  This representation first appeared in the concept of bit threads~\cite{Freedman:2016zud}, but we will continue the discussion within the broader framework of entanglement threads~\cite{Lin:2025yko}. Note that in this temporary schematic diagram, we view the threads as roughly parallel and uniformly distributed. However, as we reviewed earlier, the entanglement thread configuration can be refined further—and this refinement will be the main focus of this work.

In this paper, we consider only the planar BTZ black hole~\cite{Banados:1992wn, Banados:1992gq }. Our method can be directly generalized to the spherical BTZ black hole, but due to technical and conceptual subtleties, we leave that for another paper. The three-dimensional planar BTZ black hole, also called the BTZ black string, has the metric:
\begin{equation}
ds^2 = -\left(\frac{\Upsilon^2 - \Upsilon_+^2}{L^2}\right)d\iota^2 + \left(\frac{\Upsilon^2 - \Upsilon_+^2}{L^2}\right)^{-1}d\Upsilon^2 + \Upsilon^2 d\Phi^2,\quad -\infty < \Phi < +\infty \label{pla}.
\end{equation}
Here, $L$ is the AdS radius, and $\Phi$ labels the infinite line (denoted as $U$) at the asymptotic boundary, where the dual finite-temperature CFT conformally resides. The horizon is located at $\Upsilon_+$, and its corresponding inverse temperature $\beta_{\text{BH}}$ is related via
\begin{equation}
\beta_{\text{BH}} = \frac{2\pi L^2}{\Upsilon_+} \label{tem1}.
\end{equation}
Calculations from field theory and holography tell us that the von Neumann entropy of a subsystem $A$ of size $d = R\Delta \Phi$ in this finite-temperature state of the dual CFT is~\cite{Calabrese:2004eu, Ryu:2006bv}:
\begin{equation}
S_A = \frac{c}{3} \log\left(\frac{\beta}{\pi \varepsilon} \sinh\left(\frac{\pi d}{\beta}\right)\right) \label{btzen},
\end{equation}
where $\varepsilon$ is the UV cutoff in the field theory, and $c$ is the central charge. Using the holographic relation \eqref{cl}, this formula can be translated into holographic geometric language, where $\varepsilon$ corresponds to a radial cutoff, and $4G_N S_A$ gives the geodesic length of the RT surface corresponding to $A$. Interestingly, when the size $d$ of $A$ becomes sufficiently large, the schematic shape of the RT surface (see Fig.\ref{22.2}) seems to have a part that is nearly parallel to the horizon~\cite{Ryu:2006bv,Ryu:2006ef}. Naturally, from the thread picture viewpoint, we are led to the following intuition:

\begin{figure}
    \centering
    \includegraphics[scale=1]{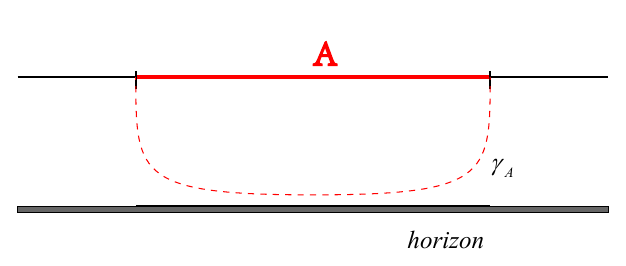}
    \caption{ When the size $d$ of $A$ becomes sufficiently large, the RT surface related to $A$ has a part that is nearly parallel to the horizon. }
    \label{22.2}
\end{figure}

1.	The above experience from wormhole/thermofield double state correspondence tells us that thermal entropy can be represented as the number of threads from the horizon surface to the boundary spatial slice.

2. The area of the RT surface $\gamma_A$ has a portion proportional to the horizon area, which in turn is proportional to $d$.

Thus, it is natural to decompose $S_A$ into a ``horizon contribution” and a contribution from the remaining part of the CFT system. 
Note that in the context of bit threads, this natural idea has indeed been applied to the analysis, see~\cite{Kudler-Flam:2019oru, Agon:2018lwq, Mintchev:2022fcp, Caggioli:2024uza}. \footnote{Such decomposition considerations also appeared early in tensor network contexts, e.g., see~\cite{Swingle:2009bg, Molina-Vilaplana:2011ydi, Matsueda:2012xm, Molina-Vilaplana:2014mna, Bao:2015uaa}.}. However, as we will discuss in Section~\ref{sec41}, the approach taken in this paper will lead to more refined results that are distinct from those of previous studies. Nevertheless, following the systematic scheme of representing von Neumann entropy via thread picture as in the introduction, we now propose a partial entanglement entropy~\cite{Vidal:2014aal} prescription adapted to the planar BTZ black hole. As shown in the FIG.~\ref{22.2}, we again divide the holographic quantum system $U$ into $N$ disjoint elementary regions, denoted $A_1, A_2, \ldots, A_N$, \footnote{One can choose $N$ large enough so that each elementary region is very small, but we assume each region is still much larger than the Planck scale to ensure the validity of the RT formula. Also, since the BTZ black hole has an infinite boundary line, we implicitly assume an IR cutoff, which will be discussed again in Section~\ref{sec4}.} that is:
\begin{equation}
A_i \cap A_j = \emptyset,\quad \bigcup A_i = U.
\end{equation}
A natural approach is to make all $N$ elementary regions of equal size (which is feasible for the one-dimensional spatial slice of the $CFT_2$). In addition, unlike the pure AdS case, we now introduce ``elementary reference lines” in the bulk, defined as equal-$\Phi$ slices in the coordinates of \eqref{pla}. These reference lines extend from each endpoint of region $A_i$ to the horizon $\Gamma$, defining a corresponding elementary region $\sigma_i$ on the horizon, such that:
\begin{equation}
\sigma_i \cap \sigma_j = \emptyset,\quad \bigcup \sigma_i = \Sigma.
\end{equation}
Under this setup, we imagine that between any two elementary regions $A_i$ and $A_j$, there exist a number of connecting ``threads”, which can be interpreted as concrete realizations of entanglement~. For any elementary region $A_i$, each other region $A_j$ provides a certain number $F_{ij}$ of threads starting from $A_j$ and arriving at $A_i$. In the FIG.~\ref{22.2}, several such threads arriving at a region $A_s$ are depicted (green and red threads). However, unlike in the pure AdS case, there are also threads originating from the horizon and ending at $A_s$ (assumed to descend uniformly), shown in gray. 
\begin{figure}
    \centering
    \includegraphics[scale=0.76]{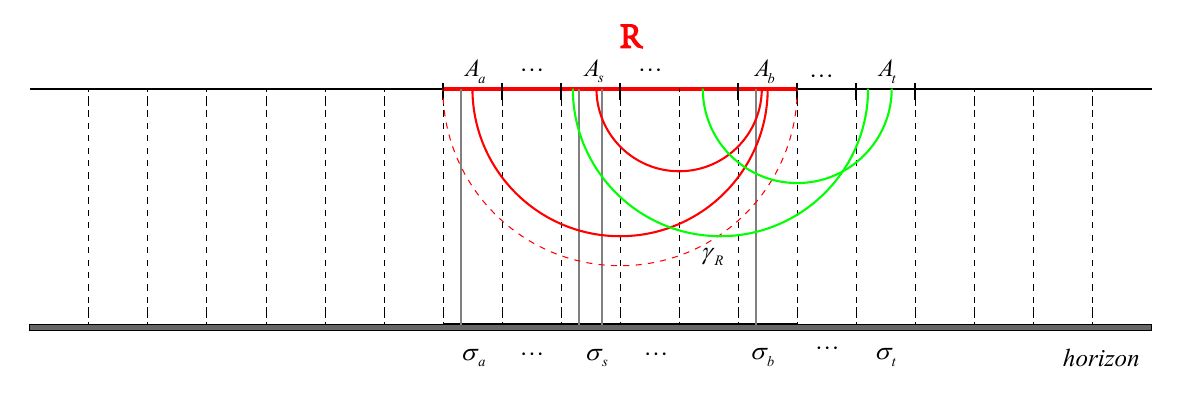}
    \caption{ The entanglement thread scheme for a holographic planar BTZ black hole setup. Some (but not all) of the entanglement threads are schematically presented. In this thread picture, to calculate the entropy of a subregion $R =  A_a \cup A_{a+1} \cup \cdots \cup A_{a+r-1}$, the threads marked by green lines (representing external entanglement) and grey lines (representing contribution from the horizon) should be counted, while those marked by red lines (representing internal entanglement) makes no contribution.}
    \label{22.3}
\end{figure}
A natural assumption is that these threads start from the horizon region $\sigma_s$ corresponding to $A_s$.\footnote{In Section~\ref{sec3}, we will revisit and justify this assumption.} Then, the von Neumann entropy of $A_s$ consists of two contributions: the number of threads connecting $A_s$ to the rest of the system, and the number of threads coming from the horizon. More generally, consider a connected composite region $R = A_{a(a+1)\cdots(a+r-1)} \equiv A_a \cup A_{a+1} \cup \cdots \cup A_{a+r-1}$, its von Neumann entropy has two parts: (i) the sum of $F_{st}$ over $A_s \in R$, $A_t \in \bar{R}$, and (ii) the number of threads flowing into $R$ from the horizon, proportional to its size. Let the size of each elementary region be $\varpi$, and let $R$ contain $r$ elementary regions, then the horizon contribution is $r\varpi\rho$, where $\rho$ is the flow density from the horizon. Thus:
\begin{equation}
S_{a(a+1)\cdots(a+r-1)} = \sum_{s,t} F_{st} + r\varpi\rho,
\quad \text{where } s \in \{a, \dots, a+r-1\},\; t \notin \{a, \dots, a+r-1\} \label{sar}.
\end{equation}
In the diagram, green and gray threads contribute to $S_{a(a+1)\cdots(a+r-1)}$. Red threads, with both ends inside $R$, represent internal entanglement and do not contribute to the entropy. As before~\cite{Lin:2021hqs}, the system of equations \eqref{sar} can be solved uniquely if we consider all connected regions $A_{a(a+1)\cdots(a+r-1)}$. There are $N(N-1)/2$ independent equations and an equal number of variables $F_{st}$. There is a convenient way to directly solve this. Rewrite \eqref{sar} as:
\begin{equation}
S_{a(a+1)\cdots(a+r-1)} - r\varpi\rho = \sum_{s,t} F_{st}, \quad \text{where } s \in \{a, \dots, a+r-1\},\; t \notin \{a, \dots, a+r-1\} \label{sar2}.
\end{equation}
Similar to equations~\eqref{equ} and \eqref{cmi}, we immediately obtain the solution to \eqref{sar2} as:
\begin{equation}
\begin{aligned}
F_{ij} &= \frac{1}{2} \left[
\left(S(\tilde{A} \cup A_i) - (l+1)\varpi\rho\right) +
\left(S(\tilde{A} \cup A_j) - (l+1)\varpi\rho\right) \right. \\
&\quad \left. - \left(S(\tilde{A}) - \varpi\rho\right) -
\left(S(A_i \cup \tilde{A} \cup A_j) - (l+2)\varpi\rho\right)
\right] \label{fijc}
\end{aligned},
\end{equation}
where we suppose $\tilde{A} = A_{(i+1)\cdots(j-1)}$ contain $l$ regions. \eqref{fijc} simplifies to:
\begin{equation}
F_{ij} = \frac{1}{2}\left[S(\tilde{A} \cup A_i) + S(\tilde{A} \cup A_j) - S(\tilde{A}) - S(A_i \cup \tilde{A} \cup A_j)\right] \equiv \frac{1}{2}I(A_i, A_j|\tilde{A}) \label{cmi2}.
\end{equation}
This is consistent in form with \eqref{cmi}. The reason is that the contributions from the horizon cancel algebraically.

Using \eqref{btzen}, we can compute explicit expressions for $F_{ij}$. Let $a_i$, $a_j$, and $\tilde{a}$ be the sizes of $A_i$, $A_j$, and $\tilde{A}$. Then:
\begin{equation}
F_{ij} = \frac{1}{2}I(A_i, A_j|\tilde{A}) = \frac{c}{6} \ln\left(\frac{
\sinh\left(\frac{\pi(a_i + \tilde{a})}{\beta}\right)
\sinh\left(\frac{\pi(a_j + \tilde{a})}{\beta}\right)}{
\sinh\left(\frac{\pi(a_i + \tilde{a} + a_j)}{\beta}\right)
\sinh\left(\frac{\pi \tilde{a}}{\beta}\right)}\right) \label{fij2}.
\end{equation}
Again, note that the expression for $F_{ij}$ is cutoff-independent.

Now, consider a region $R = A_{a(a+1)\cdots(a+r-1)}$ and one of its components $A_s$. Define the partial entanglement entropy $s_R(A_s)$ as the contribution of $A_s$ to $S_R$~\cite{Vidal:2014aal}, which should be:
\begin{equation}
s_R(A_s) = \sum_t F_{st} + \varpi\rho,\quad s \in \{a,\dots,a+r-1\},\; t \notin \{a,\dots,a+r-1\}.\label{should}
\end{equation}
Substitute \eqref{cmi2} into \eqref{should}, one obtains
\begin{equation}
s_R(A_s) = \frac{1}{2}\left[
S(L \cup A_s) + S(L \cup \bar{R}) - S(L) - S(A_s \cup L \cup \bar{R})
\right] + \varpi\rho,
\end{equation}
where $L$ is the region between $A_s$ and $\bar{R}$. More compactly:
\begin{equation}
s_R(A_s) = \frac{1}{2}I(A_s, \bar{R}|L) + \varpi\rho \label{srs}.
\end{equation}
This expression clearly shows that in the case of a finite-temperature quantum state dual to a BTZ black hole, the contribution of a subregion $A_s \subset R$ to $S_R$ consists of two parts: entanglement with the complement, and a thermal contribution. Equation~\eqref{srs} can be seen as a finite-temperature extension of the partial entanglement entropy expression using conditional mutual information proposed in~\cite{Lin:2021hqs,Lin:2022flo}. The key point is to realize that the second term in \eqref{srs} is essential. Only with it does the following equation hold:
\begin{equation}
\sum_{A_s \in R} s_R(A_s) = S_R,
\end{equation}
which is the fundamental requirement of partial entanglement entropy decompositions~\cite{Vidal:2014aal}. Again, we should add that the expression \eqref{fij2} regarding the planar BTZ black hole is not new; it can be found in reference e.g.~\cite{Czech:2015qta,Kudler-Flam:2019oru,Wen:2018whg, Asplund:2016koz, Boldis:2021snw} in various disguises. However, the discussion on the partial entanglement entropy presented here is new, and it is made possible by our understanding of the entanglement threads from the complete perspective. We will revisit this point in Section~\ref{sec4}.

\section{A More Refined Proposal: Thread Configurations from the Two-Sided Perspective}\label{sec3}

In this section, we propose a more refined entanglement thread configuration to analyze the entanglement structure of the planar BTZ black hole. To understand what we mean by ``more refined", note that in the heuristic version proposed in the previous section, we simply assumed all thread flows from the horizon are parallel. However, just as in the pure AdS/CFT vacuum correspondence, once the region is further refined, the thread flows exhibit more delicate paths in the bulk to reveal finer entanglement structures among the subdivided subsystems—these typically are no longer simple parallel threads. In fact, we know that the trajectories of these threads should be geodesics in the holographic bulk\cite{Lin:2025yko}. In the more precise version we are about to present, we also subdivide the asymptotic region on the other side of the full BTZ wormhole. Thus, we should expect a more refined thread configuration that tells us how thread flows from the elementary regions on the other side of the wormhole contribute to the entanglement entropy of subregions on this side.

\subsection{Basic Setup}\label{sec31}

Our approach simply uses the following well-known fact: the equal-time slice of the planar BTZ black hole can be obtained from a mapping of half of the Poincaré disk (see e.g.,~\cite{Asplund:2016koz}). In particular, the black hole horizon corresponds to the diameter of the Poincaré disk. From a global viewpoint, the full Poincaré disk can equivalently represent the two-sided planar BTZ black hole and the two things can be transformed into one another by a coordinate transformation. Since we already know that, at least in the pure AdS/CFT vacuum correspondence, the refined thread trajectories are exactly spacelike geodesics in the holographic bulk slice\cite{Lin:2025yko}, we can naturally construct the thread configuration in the BTZ wormhole case via this mapping. Let us now introduce the basic setup of this correspondence.

The planar BTZ black hole metric ~\eqref{pla} can be obtained from AdS$_3$ via the following coordinate transformation:
\begin{equation}
\begin{aligned}
X^0 &= \Upsilon' \cosh \Phi', \\
X^3 &= \sqrt{{\Upsilon'}^2 - L^2} \sinh \left(\frac{\iota'}{L}\right), \\
X^1 &= \sqrt{{\Upsilon'}^2 - L^2} \cosh \left(\frac{\iota'}{L}\right), \\
X^2 &= \Upsilon' \sinh \Phi'
\end{aligned} \label{tobtz}.
\end{equation}
Then, the metric~\eqref{tra} becomes:
\begin{equation}
ds^2 = -\left(\frac{{\Upsilon'}^2 - L^2}{L^2}\right) d\iota'^2 + \left(\frac{{\Upsilon'}^2 - L^2}{L^2}\right)^{-1} d\Upsilon'^2 + {\Upsilon'}^2 d\Phi'^2 \label{pla2}.
\end{equation}
To see that this metric is completely equivalent to the previous planar BTZ form \eqref{pla}, we simply apply a rescaling:
\begin{equation}
\Upsilon' = \frac{L}{\Upsilon_+} \Upsilon, \quad \iota' = \frac{\Upsilon_+}{L} \iota, \quad \Phi' = \frac{\Upsilon_+}{L} \Phi.
\end{equation}
In particular, the horizon at $\Upsilon = \Upsilon_+$ corresponds to $\Upsilon' = L$. For our purposes, another useful coordinate transformation is to reparametrize the radial coordinate of the BTZ geometry by:
\begin{equation}
\Upsilon' = L \cosh \Theta \label{repa}.
\end{equation}
Now, focusing on the equal-time slice $X^3 = 0$ (i.e., $\iota' = 0$)\footnote{Note that this equal-time slice choice is consistent with the one used in the global representation ~\eqref{glo}, so we can safely discuss everything within the same two-dimensional Poincaré disk.}, the metric~\eqref{pla2} becomes:
\begin{equation}
ds^2 = L^2 \left(d\Theta^2 + \cosh^2 \Theta\, d\Phi'^2 \right) \label{up}.
\end{equation}
Here, $\Theta$ represents the radial coordinate of the black hole, and $\Phi'$ parameterizes the $\Theta$-constant slices. From ~\eqref{tobtz}, we can see that the metric~\eqref{up} exactly covers the upper half of the Poincaré disk (i.e., the region $\Upsilon' \geq L$). We can explicitly relate the coordinates $(\Theta, \Phi')$ to the standard Poincaré disk coordinates $(x, y)$ as follows:
\begin{equation}
\begin{aligned}
x &= R \cdot \frac{\cosh \Theta \cdot \sinh \Phi'}{1 + \cosh \Theta \cdot \cosh \Phi'}, \\
y &= R \cdot \frac{\sinh \Theta}{1 + \cosh \Theta \cdot \cosh \Phi'}
\end{aligned}.
\end{equation}
Hence, the trajectory of a constant $\Theta$ surface (i.e., a constant $\Upsilon$ surface in the BTZ black hole) in the Poincaré disk is a circle given by:
\begin{equation}
x^2 + \left(y + R \coth \Theta \right)^2 = \left(R \coth \Theta \right)^2 \label{conthe}.
\end{equation}
In particular, $\Theta = 0$ corresponds precisely to the horizon $\Gamma$, i.e., $\Upsilon' = L$ or $\Upsilon = \Upsilon_+$, while $\Theta \to \infty$ corresponds to the boundary $U$ of the upper Poincaré disk. We also find the trajectory of constant $\Phi'$ in the disk. It turns out to also be a circle:
\begin{equation}
(x - R \coth \Phi')^2 + y^2 = \left(\frac{R}{\sinh \Phi'}\right)^2 \label{conphi}
\end{equation}
Equation~\eqref{conphi} defines a geodesic: in the Poincaré disk, circular arcs perpendicular to the boundary are geodesics. One can verify that~\eqref{conphi} satisfies this perpendicularity condition (whereas~\eqref{conthe} does not, and hence is not a geodesic). Moreover, since~\eqref{conphi} is a circle centered at $(R \coth \Phi', 0)$, and since the horizon $y = 0$ is its symmetry axis, the constant-$\Phi'$ curves are perpendicular to the diameter of the Poincaré disk that corresponds to the horizon. In fact, these constant-$\Phi'$ trajectories can be used to describe the ``elementary reference lines" defined in Section~\ref{sec22}, as we shall now elaborate.

\subsection{Thread Configurations from the Two-Sided Perspective}

\subsubsection{A More Refined Proposal}

\begin{figure}
    \centering
    \begin{subfigure}[b]{0.46\textwidth}
         \centering
        \includegraphics[width=\textwidth]{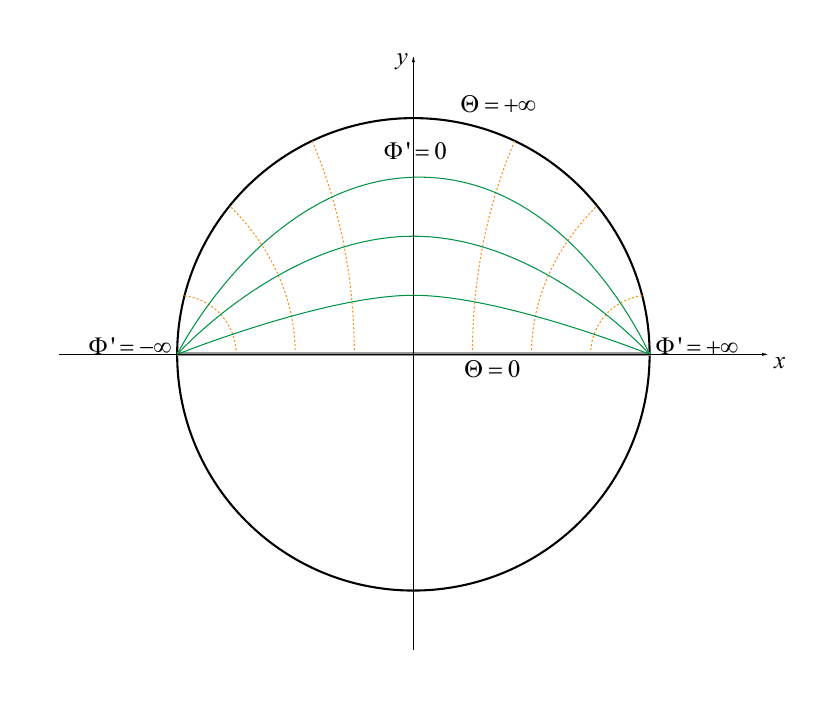}
         \caption{}
    \label{321.1a}
    \end{subfigure}
    \begin{subfigure}[b]{0.46\textwidth}
         \centering
         \includegraphics[width=\textwidth]{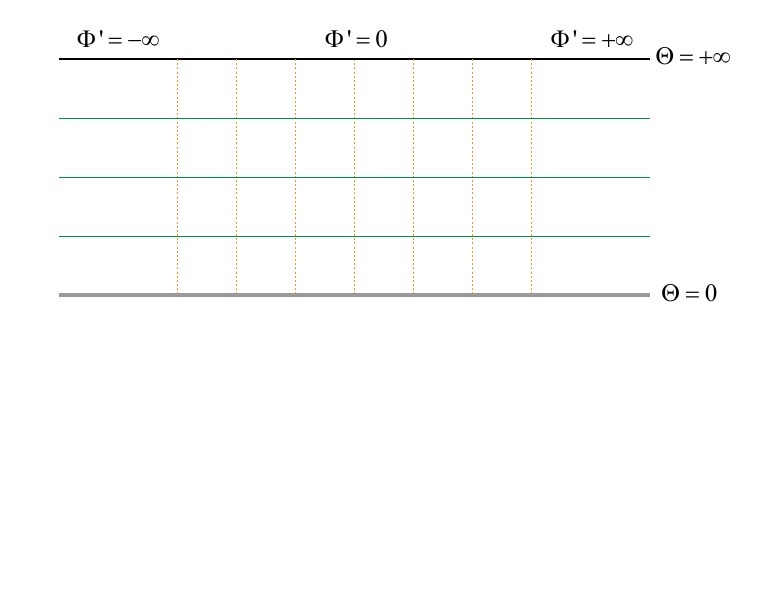}
         \caption{}
    \label{321.1b}
    \end{subfigure}
    \caption{(a) From a global perspective, the one-sided planar BTZ black hole corresponds to the upper half of the Poincaré disk. 
(b) The one-sided BTZ in the standard black hole coordinate representation $(\Theta, \Phi')$.
In both representations, we have marked the constant-$\Theta$ surfaces (green lines) and constant-$\Phi'$ surfaces (yellow dashed lines), whose trajectories are described by Eq.~\eqref{conthe} and Eq.~\eqref{conphi}, respectively. The asymptotic boundary of the black hole is represented by a solid black line, while the horizon is depicted as a thick gray line.}
    \label{321.1}
\end{figure}

From a global perspective, the BTZ black hole corresponds to only the upper half of the Poincaré disk. If we consider the two-sided BTZ black hole, it can be mapped onto the full Poincaré disk, as illustrated in Fig.~\ref{321.1}. More specifically, we analytically continue the parametrization~\eqref{repa}. Noting that the $\cosh$ function is an even symmetric function, although $\Upsilon'$ is constrained to be positive, the range of $\Theta$ can be extended from $(0,\infty)$ to $(-\infty, \infty)$, with the negative values corresponding to the lower half of the Poincaré disk. Likewise, the ``elementary reference lines” constructed earlier—i.e., the constant $\Phi'$ geodesics—can be symmetrically extended to the lower half of the disk. The trajectory equations~\eqref{conthe} and~\eqref{conphi} remain valid under this extension.

\begin{figure}
    \centering
    \begin{subfigure}[b]{0.4\textwidth}
        \centering
        \includegraphics[width=\linewidth]{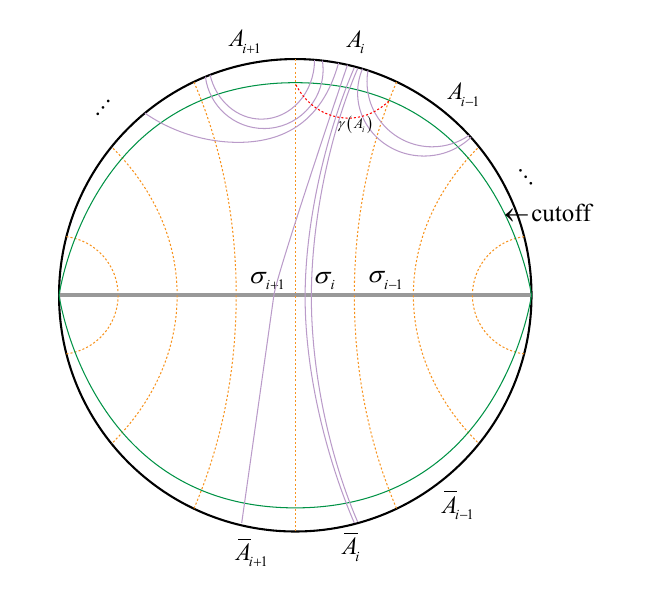}
        \caption{}
        \label{321.2a}    
    \end{subfigure}
    \hfill
    \begin{subfigure}[b]{0.49\textwidth}
        \centering
        \includegraphics[width=\linewidth]{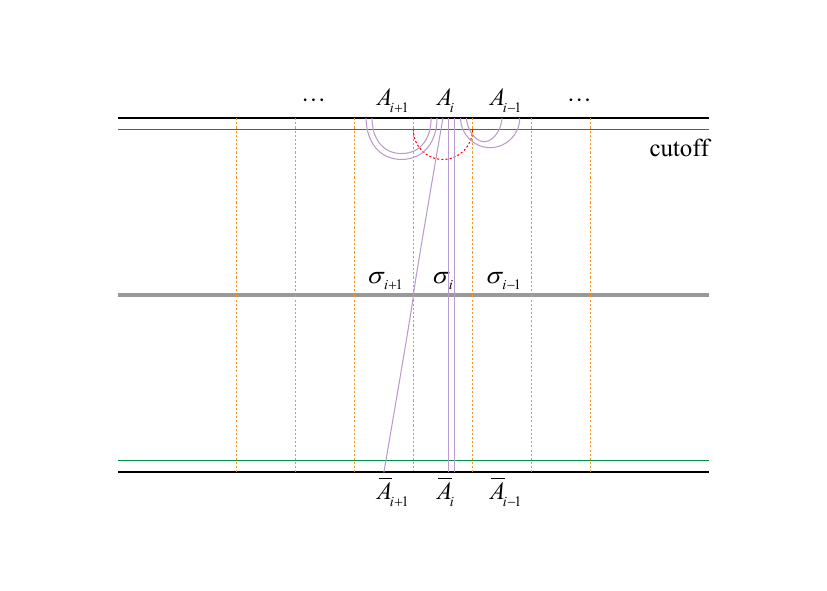}
        \caption{}
        \label{321.2b}    
    \end{subfigure}
    \caption{ The more refined entanglement thread configuration for BTZ planar black hole. Some of entanglement threads (purple curves) are schematically presented. Figure (a) shows the entanglement threads contributing to the entropy of a boundary subregion ${A_i}$ in the Poincaré disk, which has been well-understood. By translating (a) into the scenario of the two-sided BTZ black hole in Figure (b), it becomes clear that the entanglement entropy contribution of a subregion ${A_i}$ stems not only from entanglement threads originating from other elementary regions on the same side but also from entanglement threads connecting to opposite-side regions.}
    \label{321.2}
\end{figure}

Thus, as illustrated, the one-sided configuration considered in Section~\ref{sec22} can now be understood from the global, two-sided perspective. Selecting constant $\Phi'$ lines (equivalently, constant $\Phi$ lines)  with a constant spacing $\Delta \Phi' = w$, they serve as the elementary reference lines introduced in Section~\ref{sec22}. Their intersections with the spatial boundary $U$ of CFT$_1$ define elementary boundary regions $\cdots,\; A_{i-1},\; A_i,\; A_{i+1},\; \cdots$. Similarly, their intersections with the horizon $\Gamma$ define elementary regions on the horizon $\cdots,\; \sigma_{i-1},\; \sigma_i,\; \sigma_{i+1},\; \cdots$. Now, from the metric~\eqref{up}, it is clear that these elementary reference lines, spaced by $\Delta \Phi' = w$, divide the horizon $\Gamma$ (i.e., at $\Theta = 0$) into segments of equal physical length $Lw$. However, one must be cautious: since the $\Gamma$ surface is the diameter of the Poincaré disk, its physical length is divergent. Therefore, we must introduce cutoffs at both ends of the diameter (which correspond to IR cutoffs on the infinitely extended spatial line in the BTZ representation), in order for it to be divided into $N$ segments of size $Lw$. Thus, the regularized length of $\Gamma$ becomes $NLw$. Beyond Section~\ref{sec22}, these elementary reference lines are now symmetrically extended to the other side of the full slice, as shown in the FIG.~\ref{321.2}. They intersect the lower boundary of the Poincaré disk (which conformally hosts CFT$_2$), denoted by $D$, and define the corresponding ``mirror” elementary regions $\cdots,\; \bar{A}_{i-1},\; \bar{A}_i,\; \bar{A}_{i+1},\; \cdots$. 

The more refined proposal is now as follows and illustrated in FIG.~\ref{321.2}. We first look at the left panel (Poincaré disk) and consider the von Neumann entropy of an elementary region $A_i$ on the boundary $U$. What is its partial entanglement decomposition? The answer is ready: in addition to the contributions from other elementary regions $A_j$ on the same boundary $U$—which appear as the number $F_{ij}$ of threads connecting $A_j$ and $A_i$—one should also account for contributions from each elementary region $\bar{A}_j$ on the other boundary $D$. These are encoded in the number $F_{i\,\bar{j}}$ of threads connecting $\bar{A}_j$ to $A_i$. All of this is straightforward as long as we remain within FIG.~\ref{321.1}, where our understanding of entanglement threads and conditional mutual information (CMI) in pure AdS/vacuum CFT is well-developed. However, once this language is translated into the right panel (i.e., the two-sided BTZ black hole), it yields a non-trivial statement about the entanglement structure of the BTZ wormhole: for an elementary region $A_i$ on one side $U$ of the two-sided BTZ black hole, its von Neumann entropy receives contributions from two sources: one part from other elementary regions $A_j$ on the same side $U$, with contribution quantified by $F_{ij}$; another part from each elementary region $\bar{A}_j$ on the ``other world” $D$ connected through the wormhole, with contribution given by $F_{i\,\bar{j}}$.

\subsubsection{ Computation and Results}\label{sec322}

In the BTZ coordinate system $(\Theta ,\,\Phi ')$ with metric \eqref{up}, we can calculate the length of the Ryu-Takayanagi geodesic corresponding to a subsystem $ A $ of coordinate distance $\Delta \Phi '$ in the quantum system residing on the conformal boundary. The length of a geodesic anchored to the boundary in the Poincaré disk is divergent, a radial cutoff must be introduced. The subtlety lies in that, from the viewpoint of the BTZ black hole, we should take the BTZ radial cutoff surface as a large constant $\Theta$ surface $\Theta = \Theta_0$. The geodesic length between two points at $\Theta = \Theta_0$, with spatial coordinates $\Phi_1'$ and $\Phi_2'$, is given by
\begin{equation}
l(\Theta_0,\Phi_1';\,\Theta_0,\Phi_2') = 2L \cdot \ln \left(e^{\Theta_0}\sinh \frac{\Delta \Phi'}{2}\right) \label{lbtz},
\end{equation}
where $\Delta \Phi' = |\Phi_1' - \Phi_2'|$. We can check \eqref{lbtz} by a holographic argument.  The background of the holographic lower-dimensional quantum system is conformal to the boundary of the Poincaré disk. As $\Upsilon \to \infty$, for \eqref{pla}:
\begin{equation}
ds^2 \to \frac{\Upsilon^2}{R^2}\left(-d\left(\frac{R\iota}{L}\right)^2 + R^2 d\Phi^2\right) \label{asy}.
\end{equation}
Here, we introduce a spatial scale factor $R$ at the boundary, so that the physical length of a subregion of coordinate distance $\Delta \Phi'$ in the lower-dimensional quantum system is
\begin{equation}
R\Delta \Phi = \frac{RL}{\Upsilon_+}\Delta \Phi' \label{phy}.
\end{equation}
Moreover, from \eqref{asy}, the dual inverse temperature $\beta_{CFT}$ of the lower-dimensional quantum system should be the black hole temperature \eqref{tem1} multiplied by a factor $\frac{R}{L}$, provided we recall that the inverse temperature corresponds to the periodicity of the imaginary time, that is:
\begin{equation}
\beta_{CFT} = \frac{2\pi LR}{\Upsilon_+} \label{temp2}.
\end{equation}
Therefore, using \eqref{cl} and noting $e^{\Theta_0} \sim \frac{\beta}{\pi\varepsilon}$, the expression \eqref{lbtz} precisely gives the holographic entanglement entropy of a subregion of size $d = R\Delta \Phi$ in the finite temperature state of the holographic CFT dual to the BTZ black hole \eqref{btzen}, provided we observe:
\begin{equation}
\frac{\pi (R\Delta \Phi)}{\beta_{CFT}} = \pi R\frac{\Upsilon_+}{L}\Delta \Phi'\left(\frac{\Upsilon_+}{2\pi LR}\right)^{-1} = \frac{\Delta \Phi'}{2} \label{right}.
\end{equation}

\begin{figure}[htbp]
    \centering
    \begin{subfigure}{0.38\textwidth}
        \centering
        \includegraphics[width=\linewidth]{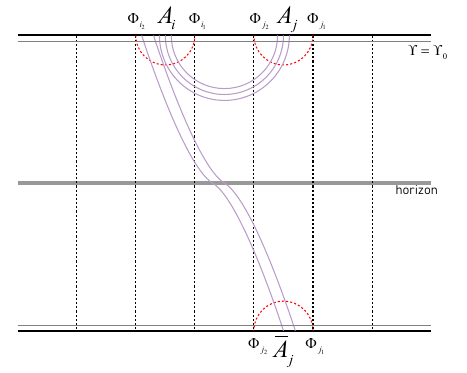}
        \caption{}
        \label{fig:322a}
    \end{subfigure}
    \begin{subfigure}{0.3\textwidth}
        \centering
        \includegraphics[width=\linewidth]{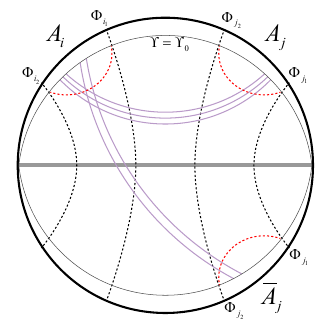}
        \caption{}
        \label{fig:322b}
    \end{subfigure}
    \\
    \begin{subfigure}{0.3\textwidth}
        \centering
        \includegraphics[width=\linewidth]{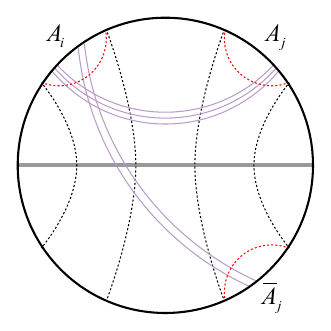}
        \caption{}
        \label{fig:322c}
    \end{subfigure}
    \begin{subfigure}{0.3\textwidth}
        \centering
        \includegraphics[width=\linewidth]{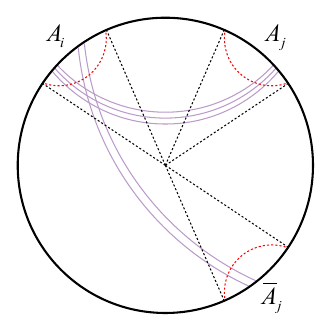}
        \caption{}
        \label{fig:322d}
    \end{subfigure}
    \begin{subfigure}{0.3\textwidth}
        \centering
        \includegraphics[width=\linewidth]{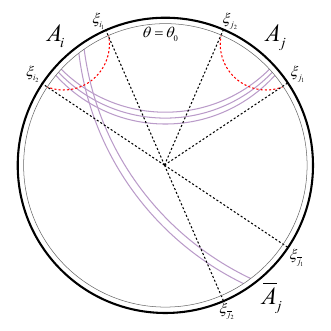}
        \caption{}
        \label{fig:322e}
    \end{subfigure}
    \caption{Illustration of the prescription for computing the entanglement thread structure of the BTZ black hole.
(a) The entanglement thread structure of the two-sided BTZ in the apparent black hole coordinates $(\Upsilon,\Phi')$. 
(b) The Poincaré disk representation of the two-sided BTZ equivalent to (a). The thread flow fluxes (i.e., half CMIs) can be calculated by first taking a radial cutoff $\Upsilon = \Upsilon_0$, or equivalently, $\Theta = \Theta_0$ to regularize the lengths of RT geodesics (red dash lines).
(c) The half CMIs are cutoff-independent.
(d) Switch back to the global coordinate system $(\theta,\,\xi)$ of the Poincaré disk.
(e) The thread flow fluxes (half CMIs) now can be equivalently calculated by adopting another kind of cutoff surface, i.e., $\theta = \theta_0$, and the results are already well-known (i.e., \eqref{fij1}).}
 \label{fig:322}
\end{figure}

On the other hand, \eqref{lbtz} and \eqref{lglo} should also be related, since as the cutoff surface tends to infinity, the surface $\theta = \theta_0$ and $\Theta = \Theta_0$ both asymptotically coincide with the holographic boundary $U$. Note that at $\Theta \to +\infty$ (i.e., on the $U$ surface), we simply have~\footnote{for $\xi \in (\pi,\,2\pi)$, it becomes $\tan \frac{\xi}{2} = -e^{-\Phi'}$.}
\begin{equation}
\tan \frac{\xi}{2} = e^{-\Phi'},\quad \xi \in (0,\,\pi) \label{equi}.
\end{equation}
This connects the global coordinates of AdS with the boundary spatial coordinates in the single-sided BTZ representation. Using \eqref{equi}, we indeed find:
\begin{equation}
\sinh \left(\frac{\Delta \Phi'}{2}\right) = \frac{1}{\sqrt{\sin \xi_1 \sin \xi_2}} \sin \left(\frac{\Delta \xi}{2}\right) \label{tran1},
\end{equation}
or
\begin{equation}
\sin \left(\frac{\Delta \xi}{2}\right) = \frac{1}{\sqrt{\cosh \Phi_1' \cosh \Phi_2'}} \sinh \left(\frac{\Delta \Phi'}{2}\right) \label{tran2}.
\end{equation}
Thus, \eqref{lbtz} and \eqref{lglo} can be transformed into each other via \eqref{tran1} and \eqref{tran2}, though the cutoff-dependent factors may differ subtly. The reason is simple: a uniform cutoff in one coordinate system appears non-uniform in the other, i.e., it depends on the angular coordinate. This issue has been recognized in~\cite{Boldis:2021snw}, and as pointed out therein, the specific choice of cutoff is irrelevant when computing conditional mutual information. In other words, the result of conditional mutual information does not depend on the specific cutoff scheme because the cutoff-dependent factors always cancel out. This provides us with a shortcut to compute the entanglement thread configuration in the BTZ black hole, as illustrated in Fig.~\ref{fig:322}.

According to Fig.~\ref{fig:322a}, we first compute the number of entanglement threads (i.e., half CMI) between two elementary regions $A_i$ and $A_j$ on the same asymptotic boundary $U$ of the BTZ black hole. This result was directly obtained in Section~\ref{sec22} from the field theory expression \eqref{fij2}. Now we compute again in the bulk using the geodesic length with cutoff, verifying that the result is cutoff-independent. Specifically, there are two cutoff schemes to compute the number of entanglement threads between $A_i$ and $A_j$. As shown in Fig~\ref{fig:322}, we can directly use the BTZ $(\Upsilon,\,\Phi')$ coordinates, or equivalently, $(\Theta,\,\Phi')$. Suppose the endpoints of the regions $A_i$ and $A_j$ in $U$ are $\Phi_{i_1}'$, $\Phi_{i_2}'$ and $\Phi_{j_1}'$, $\Phi_{j_2}'$, respectively. Their physical lengths from the field theory perspective are given by \eqref{phy}. On the bulk side, two RT geodesics corresponding to $A_i$ and $A_j$ respectively can be extended from endpoints $\Phi_{i_1}'$, $\Phi_{i_2}'$ and $\Phi_{j_1}'$, $\Phi_{j_2}'$. As shown in Figs.\ref{fig:322a} and \ref{fig:322b}, from the BTZ black hole perspective, these RT geodesics must be regularized by taking a radial cutoff $\Upsilon = \Upsilon_0$, or equivalently, $\Theta = \Theta_0$, to obtain finite lengths. Therefore, we use the geodesic length formula from \eqref{lbtz} to holographically compute half-CMI, yielding:
\begin{equation}
F_{ij} = \frac{1}{2}I(A_i,A_j|\tilde A) = \frac{c}{6} \ln \left[ \frac{\sinh\left(\frac{|\Phi_{i_1}' - \Phi_{j_1}'|}{2}\right)\sinh\left(\frac{|\Phi_{i_2}' - \Phi_{j_2}'|}{2}\right)}{\sinh\left(\frac{|\Phi_{i_2}' - \Phi_{j_1}'|}{2}\right)\sinh\left(\frac{|\Phi_{i_1}' - \Phi_{j_2}'|}{2}\right)} \right] \label{usebtz}.
\end{equation}
As expected, the cutoff dependence in \eqref{lbtz} is canceled, and according to \eqref{right}, this result exactly matches that \eqref{fij2} in Section~\ref{sec22}. Alternatively, the computation shown in Fig.~\ref{fig:322c} can be transformed into the one shown in Fig.~\ref{fig:322d}, since both consider the problem of geodesics anchored at the boundary in the Poincaré disk. Fig.~\ref{fig:322e} switches back to the global coordinate system $(\theta,\,\xi)$ of the Poincaré disk. In this case, the endpoints of $A_i$ and $A_j$ are mapped to $\xi_{i_1}, \xi_{i_2}$ and $\xi_{j_1}, \xi_{j_2}$, respectively. As reviewed in Sec\ref{sec21}, one can adopt a cutoff surface $\theta = \theta_0$ to handle this problem, and then use \eqref{lglo} to compute the result \eqref{fij1}, which is again cutoff-independent. Using the previous expressions \eqref{equi}, \eqref{tran1}, and \eqref{tran2}, one can verify that the results from the two different cutoff schemes are indeed consistent, i.e., \eqref{usebtz} is fully equivalent to \eqref{fij1}.

\begin{figure}
    \centering
    \includegraphics[width=0.6\linewidth]{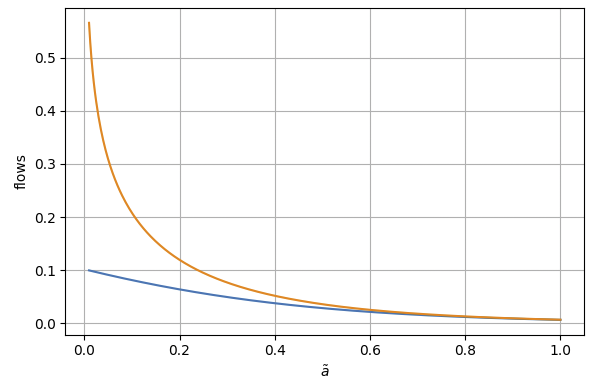}
    \caption{For a specified region $A_i$, considering the flux functions of entanglement threads arriving at $A_i$ from a region $A_j$ on the same side at a distance $\tilde{a}$ and its mirror region $\bar{A}_j$ on the opposite side (denoted as ${F_{ij}}$ (yellow curve) and $F_{i\,\bar{j}}$ (blue curve), respectively), it can be observed that the value of ${F_{ij}}$ is generally greater than that of $F_{i\,\bar{j}}$.}
    \label{plot1}
\end{figure}

Taking advantage of this equivalence, we can compute how many entanglement threads reach $A_i$ from the mirror elementary region $\bar{A}_j$, which is located on the other asymptotic boundary $D$ of the two-sided BTZ black hole. This offers a more refined partial entanglement entropy decomposition\cite{Vidal:2014aal} from the global two-sided BTZ perspective. In the analytically continued global coordinate system $(\Theta,\,\Phi')$, an elementary region $\bar{A}_j$ has endpoints with the same $\Phi'$ coordinate as its mirror image $A_j$ on the $U$ side. However, for clarity, we can also denote them by $\bar{\Phi}'_{j_1}$ and $\bar{\Phi}'_{j_2}$. Utilizing the cutoff-independence of CMI, we can similarly switch the computation to the $(\theta,\,\xi)$ coordinate system, as illustrated in Figs.~\ref{fig:322d} and ~\ref{fig:322e}. Accordingly, the angular coordinates of the endpoints of $A_i$ and $\bar{A}_j$ become $\xi_{i_1}, \xi_{i_2}$ and $\bar{\xi}_{j_1}, \bar{\xi}_{j_2}$, respectively, as shown in Figs.~~\ref{fig:322d} and ~\ref{fig:322e}. They satisfy:
\begin{equation}
\begin{aligned}
\xi_{\bar{j}_1} &= 2\pi - \xi_{j_1}, \\
\xi_{\bar{j}_2} &= 2\pi - \xi_{j_2}
\end{aligned}.
\end{equation}
Then by \eqref{fij1} we directly have:
\begin{equation}
\begin{aligned}
F_{i\,\bar{j}} &= \frac{1}{2}I(A_i,\bar{A}_j|\tilde A) \\
&= \frac{c}{6}\ln \left[\frac{\sin \left(\frac{|\xi_{\bar{j}_2} - \xi_{i_1}|}{2}\right) \sin \left(\frac{|\xi_{\bar{j}_1} - \xi_{i_2}|}{2}\right)}{\sin \left(\frac{|\xi_{\bar{j}_2} - \xi_{i_2}|}{2}\right) \sin \left(\frac{|\xi_{\bar{j}_1} - \xi_{i_1}|}{2}\right)}\right] \\
&= \frac{c}{6}\ln \left[\frac{\sin \left(\frac{|\xi_{j_2} + \xi_{i_1}|}{2}\right) \sin \left(\frac{|\xi_{j_1} + \xi_{i_2}|}{2}\right)}{\sin \left(\frac{|\xi_{j_2} + \xi_{i_2}|}{2}\right) \sin \left(\frac{|\xi_{j_1} + \xi_{i_1}|}{2}\right)}\right]
\end{aligned} \label{fijbar}.
\end{equation}
Next, we switch back to the coordinate system of the single-sided BTZ black hole. From \eqref{equi}, one can show:
\begin{equation}
\sin \left(\frac{\xi_1 + \xi_2}{2}\right) = \frac{1}{\sqrt{\cosh \Phi_1' \cosh \Phi_2'}} \cosh \left(\frac{|\Phi_1' - \Phi_2'|}{2}\right) \label{tran3}.
\end{equation}
Therefore, \eqref{fijbar} can be transformed into:
\begin{equation}
F_{i\,\bar{j}} = \frac{1}{2}I(A_i,\bar{A}_j|\tilde A) = \frac{c}{6}\ln \left[\frac{\cosh \left(\frac{|\Phi_{j_2}' - \Phi_{i_1}'|}{2}\right) \cosh \left(\frac{|\Phi_{j_1}' - \Phi_{i_2}'|}{2}\right)}{\cosh \left(\frac{|\Phi_{j_2}' - \Phi_{i_2}'|}{2}\right) \cosh \left(\frac{|\Phi_{j_1}' - \Phi_{i_1}'|}{2}\right)}\right] \label{fijbar2}.
\end{equation}
Using \eqref{right}, this can be rewritten into a more practical form on the field theory side. Denote the size of $A_i$ by $a_i$, the size of the mirror of $\bar{A}_j$ (or $ {A}_j$) by $a_j$, and the size of the region between $A_i$ and $A_j$ by $\tilde{a}$. Then we obtain a formula parallel to \eqref{fij2}:
\begin{equation}
F_{i\,\bar{j}} = \frac{c}{6}\ln \left[\frac{\cosh \left(\frac{\pi(a_i + \tilde{a} + a_j)}{\beta}\right)\cosh \left(\frac{\pi \tilde{a}}{\beta}\right)}{\cosh \left(\frac{\pi(a_i + \tilde{a})}{\beta}\right)\cosh \left(\frac{\pi(a_j + \tilde{a})}{\beta}\right)}\right] \label{nice}.
\end{equation}
This is a rather elegant form. To visually compare the relative magnitudes of the flux of entanglement threads from the same side and that from the opposite side, we present function curves for both \eqref{fij2} (yellow curve) and \eqref{nice} (blue curve) in Fig.~\ref{plot1} (where we fix the size of each elementary region as a constant, with the variable taken to be ${\tilde a}$).

Now, we compare and connect the fine-grained entanglement thread structure depicted here to the simpler configuration proposed in Section~\ref{sec22}. In that preliminary scheme, we had a rough expectation (see Eq. \eqref{srs}): for the entanglement entropy of a subregion $A_i$ in $U$, the contribution from the horizon (i.e., from the other side $D$ of the wormhole) gives a component proportional to the size of the region $A_i$, denoted by $\varpi \rho$, where $\varpi$ corresponds to the $a_i$ in our current Eq. \eqref{nice}. What we have done in this section can be understood as a more refined division of region $D$, leading to a more detailed description of the entanglement thread structure. However, the contribution of side $D$ to the entanglement entropy of subregions in side $U$ should not depend on how finely we divide $D$. In other words, after subdividing $D$, the total contribution from $D$ to $S(A_i)$ should still match the earlier coarse-grained analysis. Therefore, it should be possible to prove that summing or integrating Eq. \eqref{nice} over all $\bar{A}_j$ in region $D$ yields $\varpi \rho$. Indeed, one can prove that for a subregion $A_i$ in the upper half $U$, and its associated $\sigma_i$, the total entropy flow contribution from the lower half $D$ to $A_i$ is exactly given by the area of $\sigma_i$, that is:
\begin{equation}
a_i \rho = \frac{\text{Area}(\sigma_i)}{4G_N} = \frac{L\Delta \Phi'}{4G_N} \label{arho}.
\end{equation}
From \eqref{phy}, since $a_i = R\Delta \Phi = \frac{RL}{\Upsilon_+} \Delta \Phi'$, we obtain:
\begin{equation}
\rho = \frac{\Upsilon_+}{4G_N R}.
\end{equation}
Using \eqref{cl} and \eqref{temp2}, we find:
\begin{equation}
\rho = \frac{\pi c}{3\beta_{CFT}},
\end{equation}
which is precisely the standard thermal entropy density in the literature~\cite{Calabrese:2004eu, Ryu:2006bv}. The proof of \eqref{arho} is placed in Appendix A.

\section{Connections with previous work}\label{sec4}

As we mentioned in the background introduction, the utilization of the intuitive picture of thread configurations to describe the entanglement structure of holographic systems has been studied in different ways on multiple levels. These studies sometimes refer to essentially the same content under different names, while sometimes involve subtle conceptual differences. In this section, we specifically discuss the connections and distinctions between our work and other previous works. Following the terminology used in\cite{Lin:2025yko}, we refer to the threads studied in our work as \textbf{entanglement threads}.

\subsection{ Connection with Bit Threads}\label{sec41}

Since the thread picture we provide for the BTZ planar black hole is very similar to the bit thread picture for the BTZ planar black hole discussed in~\cite{Agon:2018lwq,Kudler-Flam:2019oru, Mintchev:2022fcp, Caggioli:2024uza}, it is necessary to emphasize the differences and connections between the two.

\begin{figure}
    \centering
    \includegraphics[width=1\linewidth]{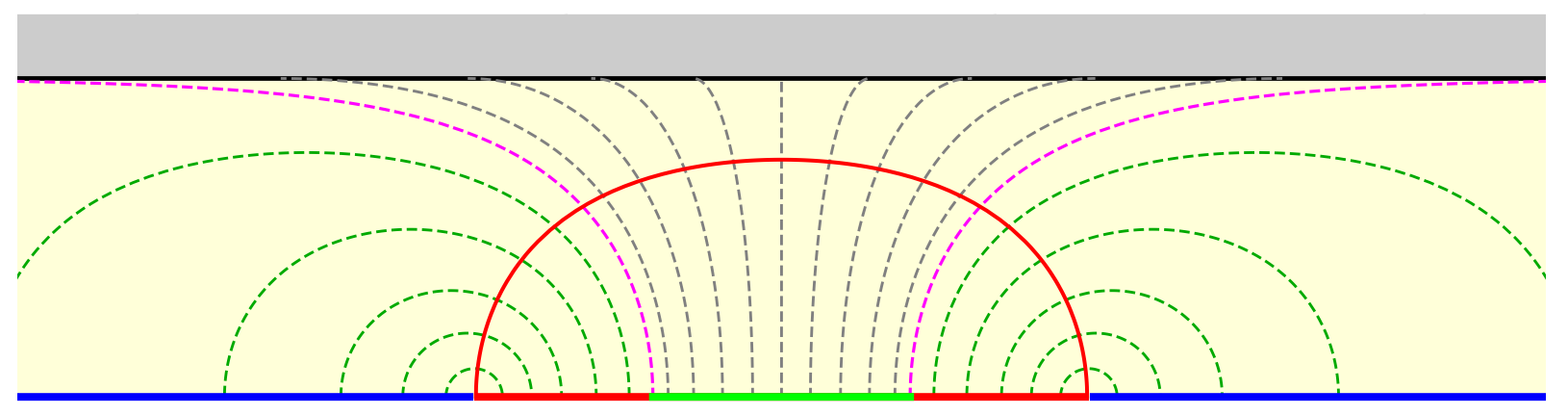}
    \caption{This picture illustrating \textbf{bit threads} is copied from~\cite{ Mintchev:2022fcp } for ease of comparison. It illustrates the bit thread configuration for a boundary subregion $ A $ (the union of the green and red segments). }
    \label{fig411}
\end{figure}

First, from a phenomenological point of view, the reader may notice that in the thread configuration of the BTZ planar black hole discussed in~\cite{Agon:2018lwq,Kudler-Flam:2019oru, Mintchev:2022fcp, Caggioli:2024uza}, all threads are constructed to be locally parallel—this is a common feature of bit threads~\cite{Freedman:2016zud,Cui:2018dyq,Headrick:2017ucz,Headrick:2022nbe}. For ease of comparison, we have copied one of their figures here (see FIG.~\ref{fig411}). It illustrates the bit thread configuration for a boundary subregion $ A $ (the union of the green and red segments). On the other hand, in our thread configuration, entanglement threads are generally intersecting. Moreover, this intersection has a physical meaning: it represents the coupling of entanglement threads through quantum gates\cite{Lin:2025yko} (as will be discussed below). In the bit thread configuration of~\cite{Agon:2018lwq,Kudler-Flam:2019oru, Mintchev:2022fcp, Caggioli:2024uza}, the locally parallel bit threads lead to the following consequence as shown in the FIG.~\ref{fig411}: contributions from the horizon surface (or from the other side of the wormhole), indicated by dashed black threads, are simply assumed to terminate in a specific green-marked subregion of $ A $. Meanwhile, the threads representing the entanglement contribution between $ A $ and its complement (the blue region on the same side) are assumed to terminate in a red-marked subregion of $ A $. In contrast, in our entanglement thread configuration, every part of $ A $ can receive entanglement contributions from the other side of the wormhole (more explicitly, by \eqref{nice}), and likewise, every part of $ A $ can receive entanglement contributions from the complementary blue region on the same side. This is the first key difference.

Of course, bit threads do not necessarily have to be locally parallel. By using the framework of \emph{multiflow}~\cite{Cui:2018dyq}, one can construct bit thread configurations with local intersections as discussed in~\cite{Headrick:2020gyq}
. Nonetheless, entanglement threads and bit threads differ in another important way. Bit thread configurations are subject to specific density bound constraints by definition. These inequality constraints imply that the bit thread configuration that computes the entanglement entropy of $ A $ is not unique. In contrast, the entanglement thread configuration is {unique}~\cite{Lin:2025yko}. This is the second key distinction between the two.

\begin{figure}[htbp]
    \centering
    \begin{subfigure}{0.3\textwidth}
        \centering
        \includegraphics[width=\linewidth]{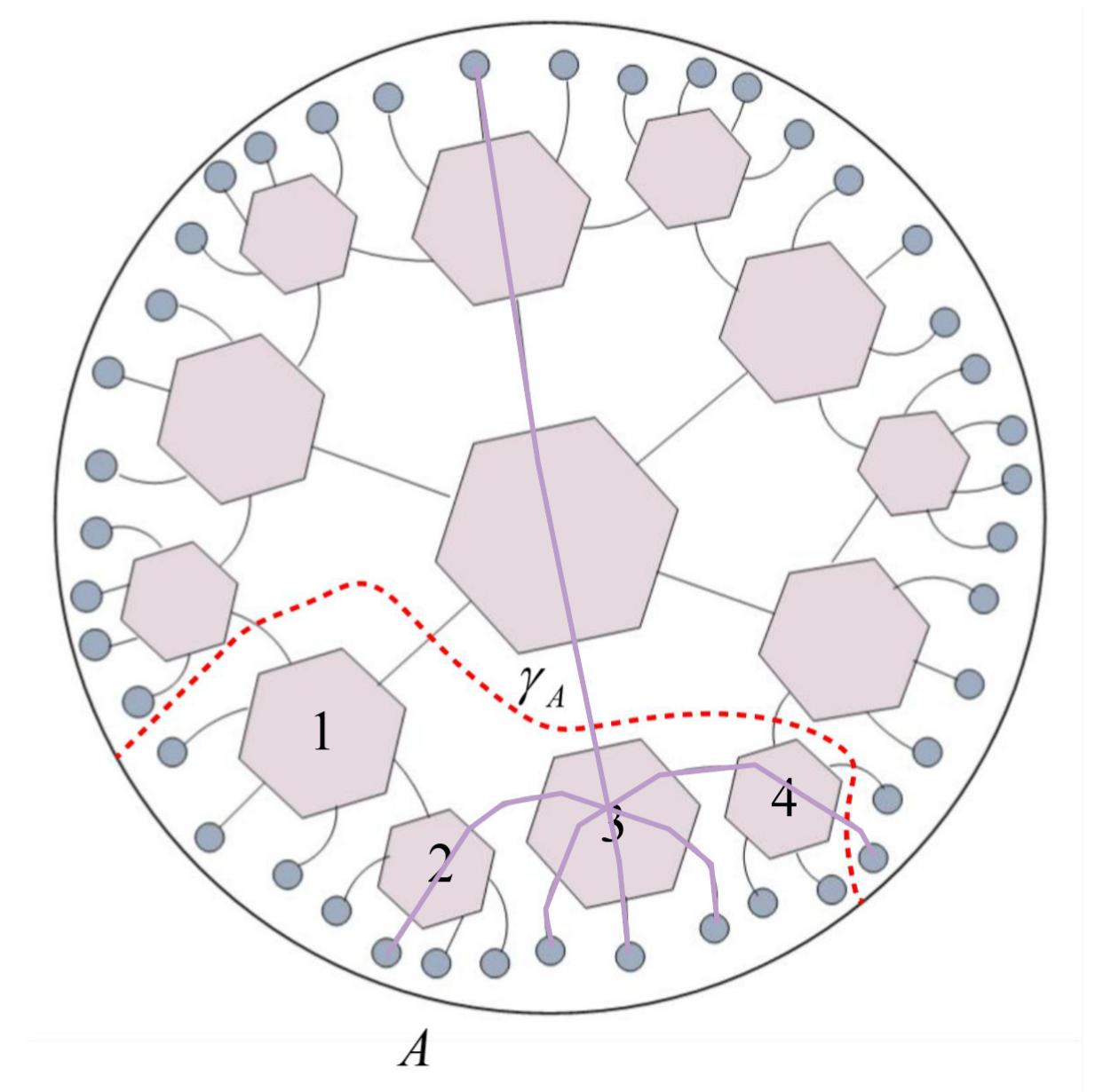}
        \caption{ }
        \label{fig:412a}
    \end{subfigure}
    \begin{subfigure}{0.3\textwidth}
        \centering
        \includegraphics[width=\linewidth]{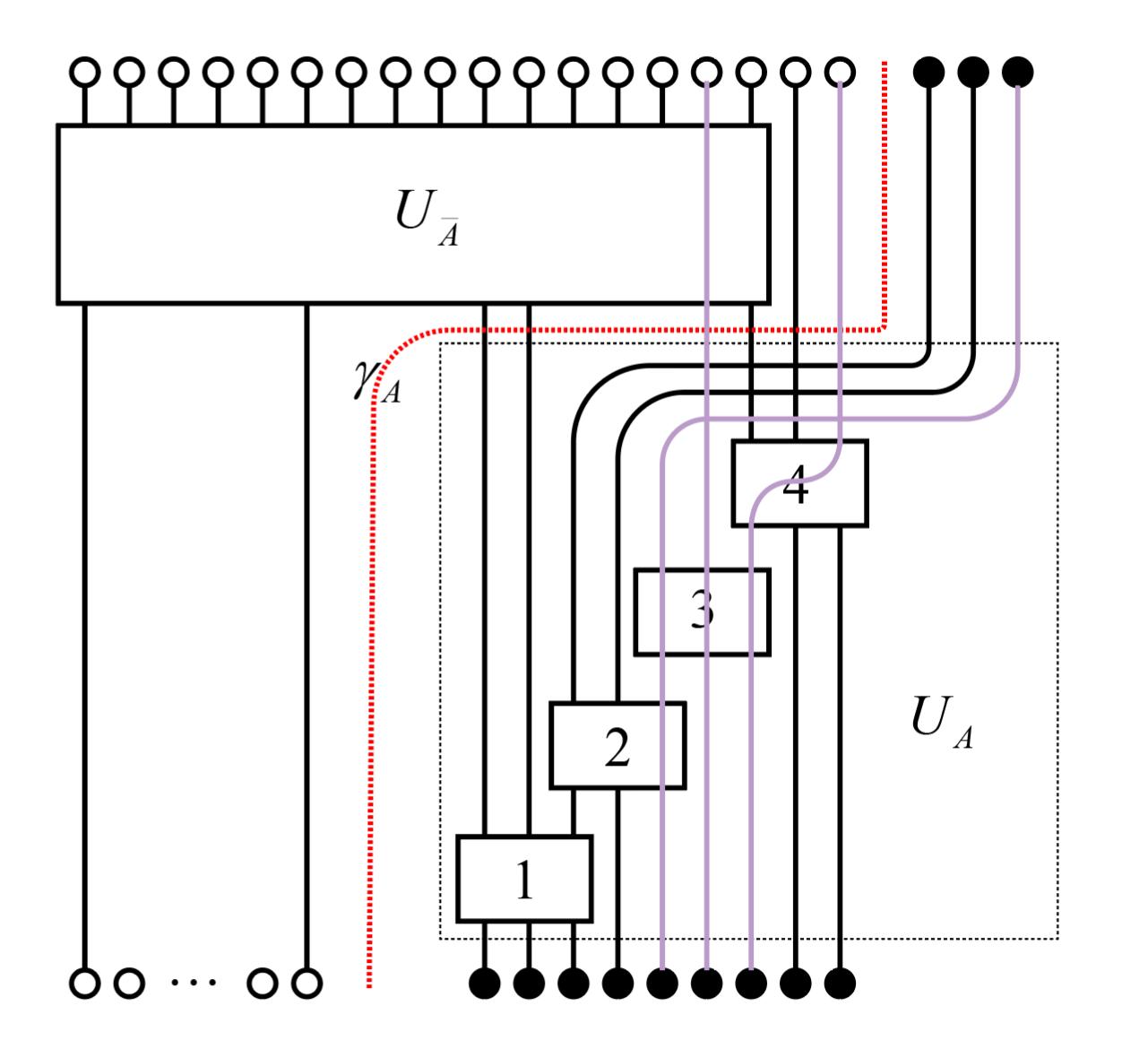}
        \caption{ }
        \label{fig:412b}
    \end{subfigure}
    \begin{subfigure}{0.34\textwidth}
        \centering
        \includegraphics[width=\linewidth]{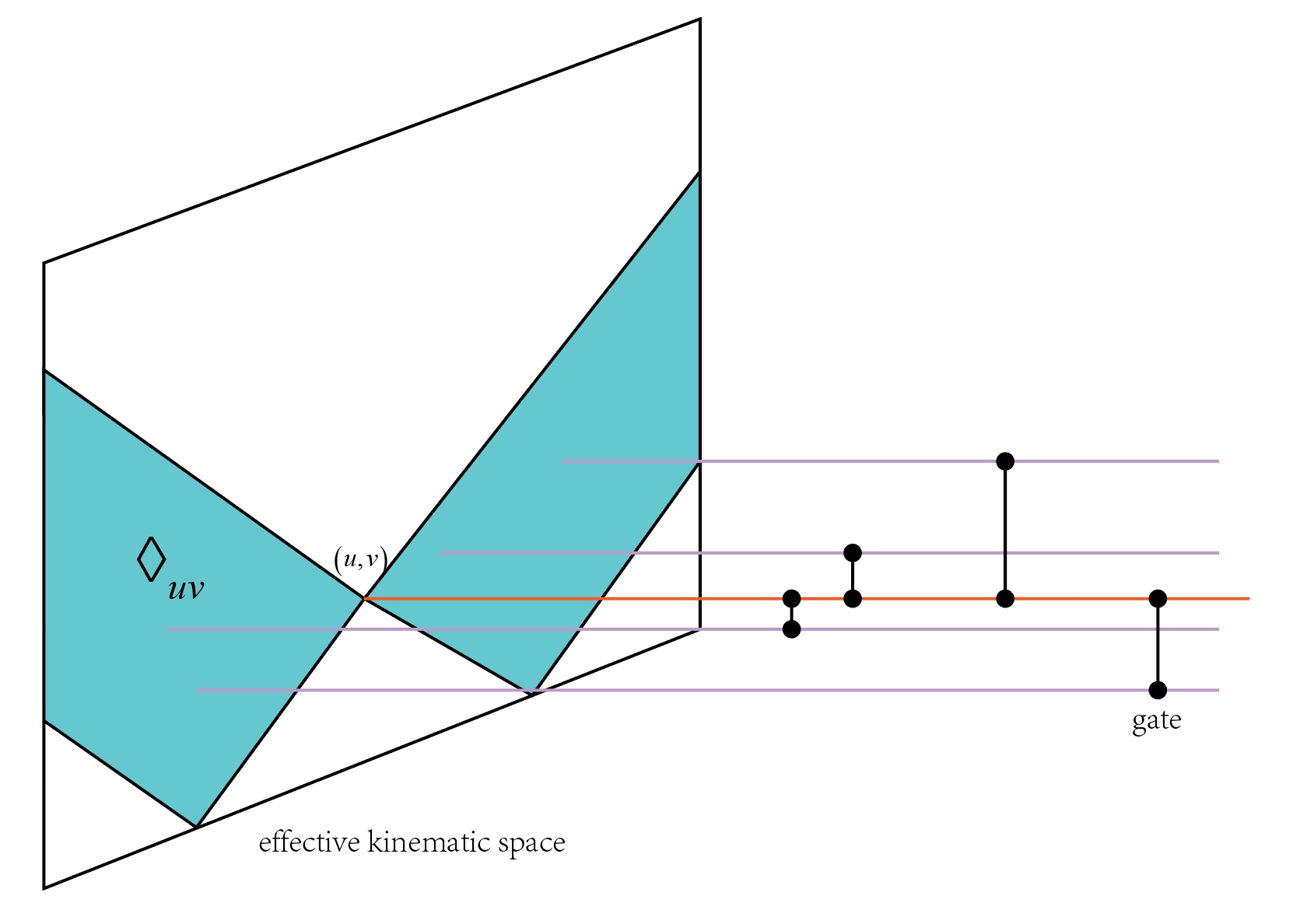}
        \caption{ }
        \label{fig:412c}
    \end{subfigure}
    \caption{ These pictures illustrating \textbf{entanglement threads} are taken from~\cite{Lin:2025yko}. For more details, please refer to the original literature. (a) The entanglement threads (purple lines) travel in the tensor network model. (b) By reorganizing the tensor network into the form of a quantum circuit, the entanglement thread configuration can be defined as the set of quantum wires. (c) Using the tools from kinematic space, each apparent wire in the circuit corresponds precisely to an entanglement thread.}
    \label{fig:412}
\end{figure}

\cite{Lin:2025yko} has proposed the following way to understand the relationship between entanglement thread configurations and bit thread configurations. As shown in the FIG.~\ref{fig:412}, the entanglement thread configuration can be defined as a set of quantum wires obtained by reorganizing the tensor network that characterizes the entanglement structure of a holographic time-slice into the form of a quantum circuit. In particular, as illustrated in Fig.~\ref{fig:412c}, when this quantum circuit is arranged according to the organization of kinematic space\cite{Czech:2015kbp,Czech:2015qta}, each apparent wire in the circuit corresponds precisely to an entanglement thread that traces a spacelike geodesic path on the bulk slice. However, in a quantum circuit, since the number of wires is conserved, we have the freedom to choose conventions for the apparent wires. For instance, Fig.~\ref{fig:412b} shows another convention, in which the apparent wires (black threads) do not coincide with the entanglement threads (purple threads). Thus, different bit thread configurations can actually be understood as different conventions for choosing the apparent wires. Nevertheless, the more subtle connections between entanglement threads and bit threads can serve as an interesting subject for further investigation. For more details discussing the differences and connections between the bit threads and entanglement threads, we recommend that readers refer to~\cite{Lin:2025yko}.

\subsection{Connections with RT Phase Transitions and Perfect-Tensor State}\label{sec42}

One advantage of analyzing holographic setups using the entanglement thread picture is that, assigning precise quantum state interpretations to these thread configurations leads to an understanding of the phase transition phenomena of RT surfaces. As it was first mentioned in~\cite{Freedman:2016zud}, as gradually adjusting the size of the subregions, although the RT surface used to compute the entanglement entropy undergoes an apparent ``phase transition’’, the corresponding thread configuration, which equivalently computes holographic entropy remains continuously varying. Moreover, the thread perspective helps explain the necessity of multipartite entanglement in the holographic duality\cite{Lin:2023orb,Lin:2023jah,Lin:2023hzs}.  

\begin{figure}[htbp]
    \centering
    \begin{subfigure}{0.485\textwidth}
        \centering
        \includegraphics[width=\linewidth]{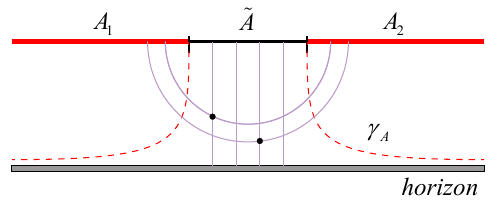}
        \caption{}
        \label{fig:42a}
    \end{subfigure}
    \begin{subfigure}{0.45\textwidth}
        \centering
        \includegraphics[width=\linewidth]{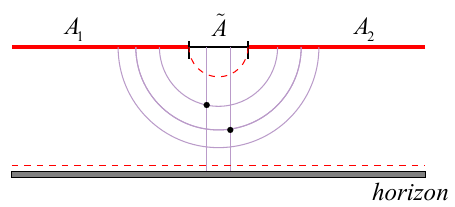}
        \caption{}
        \label{fig:42b}
    \end{subfigure}
    
    \begin{subfigure}{0.42\textwidth}
        \centering
        \includegraphics[width=\linewidth]{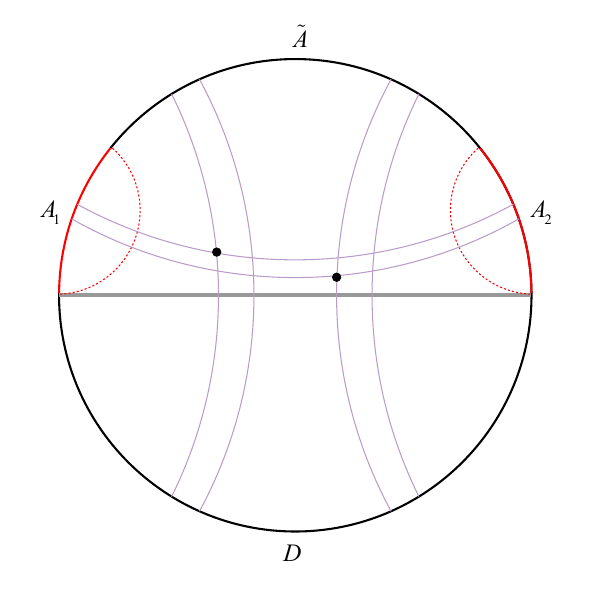}
        \caption{}
        \label{fig:42c}
    \end{subfigure}
    \begin{subfigure}{0.42\textwidth}
        \centering
        \includegraphics[width=\linewidth]{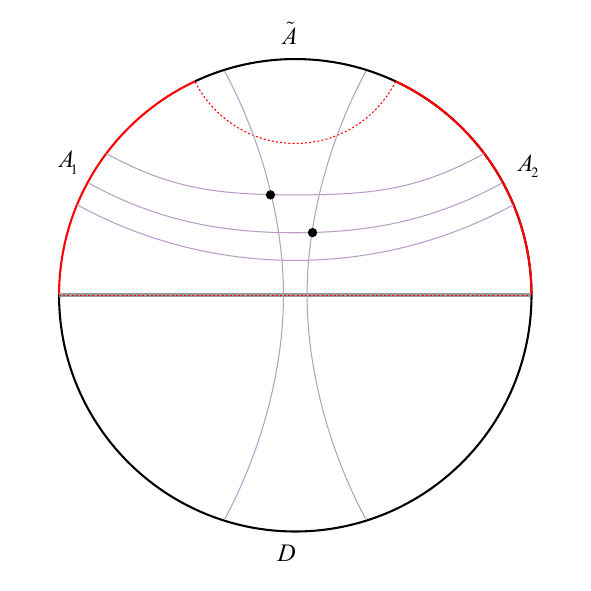}
        \caption{}
        \label{fig:42d}
    \end{subfigure}
    \begin{subfigure}{0.45\textwidth}
        \centering
        \includegraphics[width=\linewidth]{ 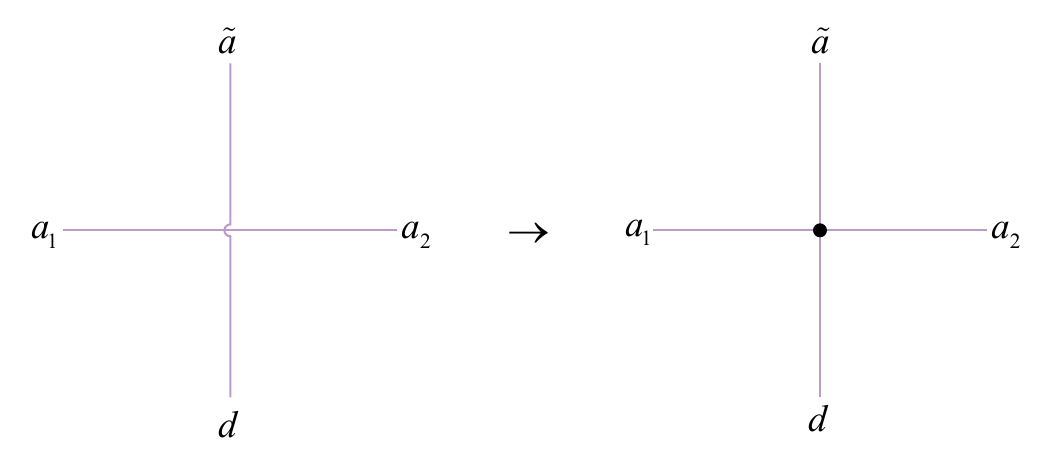 }
        \caption{}
        \label{perfect}
    \end{subfigure}
    \caption{(a), (b) For a BZT boundary region $A = A_1 \cup A_2$, depending on its size, the RT surface computing $S(A)$ has two choices, which can be determined by the thread flow competition mechanism. (c), (d) The same the competition between thread flows can be analyzed using the equivalent Poincaré disk representation, which implies the existence of perfect-type entanglement according to previous research. (e) To appropriately reflect the entanglement between $A = A_1 \cup A_2$ and $\tilde{A} \cup D$, the original decoupled quantum state \eqref{ori} corresponding to two threads $\zeta_{a_1 a_2}$ and $\zeta_{\tilde{a} d}$ should be replaced with the perfect tensor state~\eqref{per}.}
    \label{fig:42}
\end{figure}

In the case of a single-sided spherical BTZ black hole, a notable RT surface phase transition occurs when we consider the entanglement entropy of a sufficiently large boundary subregion $A$~\cite{Ryu:2006bv,Ryu:2006ef}. At this point, there are two possible shapes for the RT surface, depending on which one has the smaller area: one option is that the RT surface is connected and homologous to $A$; the other option is a union of the RT surface of the smaller complementary region $\tilde{A}$ and the entire horizon surface. Although we focus on the planar BTZ black hole in this paper, it can simulate the scenario described above. As shown in the FIG.~\ref{fig:42}, we consider the region $A = A_1 \cup A_2$ as the complement of a finite subregion $\tilde{A}$ on the asymptotic boundary $U$. Depending on the size of $\tilde{A}$, the RT surface computing $S(A)$ has two choices shown in the FIG.~\ref{fig:42}, and accordingly:
\begin{equation}
S(A) =
\begin{cases}
\displaystyle \frac{\pi c}{3 \beta_{\text{CFT}}} \cdot 2R(\Phi_{\text{IR}} - \alpha) + \frac{c}{3} \ln \left( \frac{\beta_{\text{CFT}}}{2\pi \varepsilon} \right), & \alpha \geq \frac{\beta_{\text{CFT}} \ln 2}{4\pi R} \\
\displaystyle \frac{\pi c}{3 \beta_{\text{CFT}}} \cdot 2R\Phi_{\text{IR}} + \frac{c}{3} \ln \left( \frac{\beta_{\text{CFT}}}{\pi \varepsilon} \sinh \frac{2\pi R \alpha}{\beta_{\text{CFT}}} \right), & \alpha \leq \frac{\beta_{\text{CFT}} \ln 2}{4\pi R}
\end{cases}
\label{two}.
\end{equation}

Here, we have introduced an IR cutoff, that is, we assume $\Phi \in (-\Phi_{\text{IR}},\,\Phi_{\text{IR}})$, and the $\Phi$-coordinates of the two endpoints of $\tilde{A}$ are $-\alpha$ and $\alpha$, respectively. According to our discussion, we now understand that the same result can be interpreted using the equivalent Poincaré disk representation of the whole two-sided BTZ geometry. In the Poincaré disk, this RT surface phase transition corresponds to the phenomenon that arises when calculating the entanglement entropy of the union of two disconnected boundary subregions. The IR cutoff $\Phi_{\text{IR}}$ in the former setup can be understood as the UV cutoff along the diameter geodesic in the disk. 

The entanglement thread configurations in the two representations are now seen to be equivalent, as shown in the FIG.~\ref{fig:42}. We denote the thread fluxes involved in the diagram by $\{F_{\tilde{A}A_1},\, F_{\tilde{A}A_2},\, F_{\tilde{A}D},\, F_{A_1A_2},\, F_{A_1D},\, F_{A_2D}\}$. In the Poincaré disk, the entanglement thread picture is well-developed. First of all, the RT surface phase transition can be equivalently described as competition between thread flows. Imagine gradually increasing the size of $\tilde{A}$, \eqref{two} shows that initially $S(A)$ is given by the sum of the areas of $\gamma(\tilde{A})$ and $\gamma(D)$ (corresponding to FIG.~\ref{fig:42d}). However, once the separation between $A_2$ and $A_1$ exceeds the critical value $\alpha_c = \frac{\beta_{\text{CFT}} \ln 2}{4\pi R}$, the RT surface corresponding to $S(A)$ undergoes a phase transition, that is, instead given by the sum of the areas of $\gamma(A_1)$ and $\gamma(A_2)$ (corresponding to FIG.~\ref{fig:42c}). We now employ the thread picture to understand this transition. According to the prescription in~\cite{Lin:2023orb}, we have 
\begin{equation}
\label{pres}
S(A) = \left\{ {\begin{array}{*{20}{c}}
{{F_{{A_1}\tilde A}} + {F_{{A_1}D}} + {F_{{A_2}\tilde A}} + {F_{{A_2}D}} + 2{F_{{A_1}{A_2}}},\quad \alpha  \ge {\alpha _c}}\\
{{F_{{A_1}\tilde A}} + {F_{{A_1}D}} + {F_{{A_2}\tilde A}} + {F_{{A_2}D}} +2 {F_{\tilde AD}},\quad \alpha  \le {\alpha _c}}
\end{array}} \right.\end{equation}
\eqref{pres} instructs us that when considering the RT phase transition indicated by~\eqref{two}, we only need to compare the two distinct flows $F_{A_1A_2}$ and $F_{\tilde{A}D}$.
 Initially, when the size of $\tilde{A}$ which is the relative distance between $A_2$ and $A_1$ is small, we compute the magnitudes of the six involved thread fluxes. As we increase the size of $\tilde{A}$, we specifically find that $F_{A_1A_2}$ decreases while $F_{\tilde{A}D}$ increases. In the Poincaré disk representation, denoting the size of $\tilde{A}$ in $\xi$ (or $\vartheta$) coordinates as $\Delta \xi = 2\ell$, we can explicitly compute:
\begin{equation}
\begin{aligned}
F_{A_1A_2} &= \frac{1}{2} \left[ S(A_1 \tilde{A}) + S(A_2 \tilde{A}) - S(A_1 \tilde{A} A_2) - S(\tilde{A}) \right] \\
&= \frac{c}{6} \ln \left( \frac{\sin^2\left(\frac{\pi}{4} + \frac{\ell}{2}\right)}{\sin(\ell)} \right)
\end{aligned}
\label{expl}
\end{equation}

\begin{equation}
\begin{aligned}
F_{\tilde{A}D} &= \frac{1}{2} \left[ S(\tilde{A} A_1) + S(A_1 D) - S(\tilde{A} A_1 D) - S(A_1) \right] \\
&= \frac{c}{6} \ln \left( \frac{\sin^2\left(\frac{\pi}{4} + \frac{\ell}{2}\right)}{\sin^2\left(\frac{\pi}{4} - \frac{\ell}{2}\right)} \right)
\end{aligned}
\end{equation}
If we return to the more common coordinate representation $\Phi$ of the two-sided BTZ black hole, we obtain
\begin{equation}\label{com1}
{F_{{A_1}{A_2}}} =  - \frac{c}{6}\ln (1 - {e^{ - 2\alpha '}}),
\end{equation}
\begin{equation}\label{com2}{F_{\tilde AD}} = \frac{c}{3}\alpha ',
\end{equation}
where $\alpha ' = \frac{{2\pi R\alpha }}{{{\beta _{CFT}}}}$.
From ~\eqref{com1} and~\eqref{com2}, it follows that when
\begin{equation}
\alpha' = \frac{\ln 2}{2}
\end{equation}
—which is consistent with~\eqref{two}—the two flows become equal. In this critical situation, we have
\begin{equation}
F_{A_1 A_2} = F_{\tilde{A} D} = \frac{c}{6} \ln 2.
\end{equation}
The key point, as emphasized in the original work on bit threads, is that although a distinct jump emerges from the perspective of the dual geometric surface when considering the entanglement entropy of disconnected regions, the transition remains smooth and continuous when viewed from the perspective of threads or flows.

We would like to emphasize that, in contrast to the original work on bit threads~\cite{Freedman:2016zud}, our analysis via entanglement threads more explicitly reveals that such a jump is essentially the result of a competition in magnitude between the two flows $F_{A_1 A_2}$ and $F_{\tilde{A} D}$. Furthermore, entanglement threads are inherently equipped with quantum states—as seen in Section~\ref{sec41}—they can be interpreted in the context of a quantum circuit. This provides an opportunity to understand the jump phenomenon from the perspective of specific quantum states. In fact, this jump phenomenon implies the necessity of multipartite entanglement in the holographic dual. This has been elaborately discussed in~\cite{Lin:2023orb,Lin:2023jah,Lin:2025yko}. We interpret the ideas as follows.

Suppose we begin with a trivial quantum circuit in which all wires are decoupled -- that is, each wire experiences only a trivial identity single-qudit quantum gate. In the language of tensor networks, this means that each entanglement thread (i.e., wire) connects a pair of qudits in a Bell state. Explicitly, for each entanglement thread $\varsigma$, the quantum state shared between the two qudits $x$ and $y$ at its ends is
\begin{equation}
\left| \varsigma_{xy} \right\rangle = \frac{1}{\sqrt{d}} \left( \left| 0_x 0_y \right\rangle + \left| 1_x 1_y \right\rangle + \cdots + \left| d_x d_y \right\rangle \right).
\end{equation}
In this picture, the RT surface acts as a cut in the tensor network model, and the number of legs (i.e., threads) it severs gives the entanglement entropy in the tensor network state. However, the analysis of the entanglement entropy between $A = A_1 \cup A_2$ and $\tilde{A} \cup D$ will reveal that bipartite entanglement alone is insufficient. As seen in ~\eqref{pres}, phenomenologically both the flow $F_{A_1 A_2}$ and the flow $F_{\tilde{A} D}$ can contribute (each with a factor of 2) to $S(A)$, which characterizes the entanglement between $A_1 \cup A_2$ and $\tilde{A} \cup D$. This yields an entangled state in which the contributions of both flows to this entanglement are understandable. To achieve this, the type of entanglement associated with each thread must be modified. In the language of quantum circuits, this means introducing non-trivial two-qudit quantum gates that couple different wires. Let the qudits located in $A_1$, $A_2$, $\tilde{A}$, and $D$ be denoted as $a_1$, $a_2$, $\tilde{a}$, and $d$, respectively. Let the thread connecting $a_1$ and $a_2$ be $\zeta_{a_1 a_2}$, and the one connecting $\tilde{a}$ and $d$ be $\zeta_{\tilde{a} d}$. Phenomenologically, $\zeta_{a_1 a_2}$ crosses the RT surfaces $\gamma(A_1)$ and $\gamma(A_2)$, while $\zeta_{\tilde{a} d}$ crosses $\gamma(\tilde{A})$ and $\gamma(D)$. As shown in FIG.~\ref{perfect}, the replacement rule is to substitute the original decoupled quantum state (said, with $d = 3$)
\begin{equation}
\left| \zeta_{a_1 a_2} \cup \zeta_{\tilde{a} d} \right\rangle = \frac{1}{d} \left( \left| 0_{a_1} 0_{a_2} \right\rangle + \left| 1_{a_1} 1_{a_2} \right\rangle + \left| 2_{a_1} 2_{a_2} \right\rangle \right) \otimes \left( \left| 0_{\tilde{a}} 0_d \right\rangle + \left| 1_{\tilde{a}} 1_d \right\rangle + \left| 2_{\tilde{a}} 2_d \right\rangle \right),\label{ori}
\end{equation}
with the following entangled state:
\begin{equation}
\begin{aligned}
\left| \zeta_{a_1 a_2} \cup \zeta_{\tilde{a} d} \right\rangle = \frac{1}{3} (&\left| 0_{a_1} 0_{a_2} 0_{\tilde{a}} 0_d \right\rangle + \left| 1_{a_1} 1_{a_2} 1_{\tilde{a}} 0_d \right\rangle + \left| 2_{a_1} 2_{a_2} 2_{\tilde{a}} 0_d \right\rangle \\
&+ \left| 0_{a_1} 1_{a_2} 2_{\tilde{a}} 1_d \right\rangle + \left| 1_{a_1} 2_{a_2} 0_{\tilde{a}} 1_d \right\rangle + \left| 2_{a_1} 0_{a_2} 1_{\tilde{a}} 1_d \right\rangle \\
&+ \left| 0_{a_1} 2_{a_2} 1_{\tilde{a}} 2_d \right\rangle + \left| 1_{a_1} 0_{a_2} 2_{\tilde{a}} 2_d \right\rangle + \left| 2_{a_1} 1_{a_2} 0_{\tilde{a}} 2_d \right\rangle )
\end{aligned}\label{per}.
\end{equation}
\eqref{per} is merely a concrete representation of a rank-4 perfect tensor state (or absolutely maximally entangled (AME) state)~\cite{Facchi:2008ora, Helwig:2012nha , Helwig:2013ckb, Helwig:2013qoq }. Perfect tensors are the fundamental building blocks in the well-known holographic HAPPY code~\cite{Pastawski:2015qua}, which postulates that the holographic bulk space should be constructed from entanglement types analogous to quantum error-correcting codes~\cite{Almheiri:2014lwa,Dong:2016eik}. Crucially, perfect tensor states have the property that any one qudit is maximally entangled with the remaining three, yielding an entanglement entropy of $\log 3$; and any two qudits are maximally entangled with the other two, yielding an entropy of $2 \log 3$. Note that the original state in~\eqref{ori} lacks this symmetry.

By introducing perfect entanglement, we can now illustrate the mechanism of flow competition as follows. For the disconnected region $A_1 \cup A_2$, \eqref{pres} tell us its holographic RT phase is determined by the competition between two intersecting flows: one flow $F_{A_1 A_2}$ connecting the disconnected components $A_1$ and $A_2$, and the other flow $F_{\tilde{A} D}$ connecting the remaining two disjoint complement regions $\tilde{A}$ and $D$. Each thread $\zeta_{a_1 a_2}$ from the first flow and $\zeta_{\tilde{a} d}$ from the second intersect and become entangled to form a rank-4 perfect entangled state involving $a_1$, $a_2$, $\tilde{a}$, and $d$. In the ``moving experiment’’, initially $A_2$ is close to $A_1$, so $F_{A_1 A_2} > F_{\tilde{A} D}$. Hence, all threads $\zeta_{\tilde{a} d}$ are paired with $\zeta_{a_1 a_2}$ to form $F_{\tilde{A} D}$ number of perfect tensor states. Each perfect tensor contributes $2 \log 3$ to the entanglement between $A_1 \cup A_2$ and $\tilde{A} \cup D$. Phenomenologically, this implies that $S(A)$ includes contributions not only from the flows between $A_1$, $A_2$ and their complements $\tilde{A}$, $D$, but also precisely twice the flow flux $F_{\tilde{A} D}$. This results in the RT surface $\gamma(\tilde{A}) \cup \gamma(D)$ being selected, which is crossed twice by $F_{\tilde{A} D}$ (and zero times by $F_{A_1 A_2}$). When $A_2$ is moved farther from $A_1$ such that $\tilde{A}$ exceeds the critical size, $F_{A_1 A_2} < F_{\tilde{A} D}$, and all $\zeta_{a_1 a_2}$ threads are used to form perfect entanglement with an equal number of $\zeta_{\tilde{a} d}$ threads. Thus, phenomenologically, the part of $S(A)$ contributed by $2 F_{\tilde{A} D}$ is replaced by $2 F_{A_1 A_2}$, and the corresponding RT surface becomes $\gamma(A_1) \cup \gamma(A_2)$, which is crossed twice by $F_{A_1 A_2}$ (and zero times by $F_{\tilde{A} D}$).

It is worth noting that the flow ${F_{\tilde{A}D}}$ actually characterizes the entanglement threads that flow from the other asymptotic boundary of the wormhole or the black hole horizon (i.e., $\gamma(D)$) to the subregion $\tilde{A}$ on $U$, whereas ${F_{A_1A_2}}$ corresponds to ``internal" entanglement threads with both ends lying on $U$. In summary, we are thus led to the following intriguing conclusion: for a wormhole, the entanglement threads on one side must form a perfect tensor-type entanglement with those that traverse the wormhole and come from the other side! Alternatively, from a single-sided viewpoint, this suggests that the entanglement threads that describe the entanglement structure of the holographic boundary must form a perfect entangled state with the entanglement threads coming from the horizon surface.

\subsection{Connection With Partial Entanglement Entropy Function}\label{sec43}

As reviewed earlier, the concepts of conditional mutual information and entanglement threads are also closely related to the so-called partial entanglement entropy (PEE)\cite{Vidal:2014aal}, which likewise involves a refined decomposition of the entanglement entropy. More precisely, the idea of PEE is to divide a subregion $ A $ into $ n $ degrees of freedom $\{ A_i \}$, noting that the entanglement entropy of $ A $ can be decomposed as a sum of conditional entropies:
\begin{equation}
S(A) = \frac{1}{2} \sum_{i=1}^{n} \left[ S\left( A_i \mid A_1 \cup \cdots \cup A_{i-1} \right) + S\left( A_i \mid A_{i+1} \cup \cdots \cup A_n \right) \right].
\end{equation}
Therefore, if we define\cite{Kudler-Flam:2019oru}
\begin{equation}
s_A(A_i) = \frac{1}{2} \left[ S\left( A_i \mid A_1 \cup \cdots \cup A_{i-1} \right) + S\left( A_i \mid A_{i+1} \cup \cdots \cup A_n \right) \right] 
\label{pee1},
\end{equation}
then it follows that
\begin{equation}
\sum_{A_i \in A} s_A(A_i) = S(A).
\end{equation}
Here, $ s_A(A_i) $ is called the partial entanglement entropy of $ A_i $ with respect to $ A $, meaning the contribution of the component $ A_i $ to the entanglement entropy of the region $ A $. If we denote the region to the left of $ A_i $ in $ A $ by $ A_L = A_1 \cup \cdots \cup A_{i-1} $, and the region to the right of $ A_i $ by $ A_R = A_{i+1} \cup \cdots \cup A_n $, then equation~\eqref{pee1} simplifies to\cite{Wen:2018whg}
\begin{equation}
s_A(A_i) = \frac{1}{2} \left[ S(A_i \cup A_L) + S(A_i \cup A_R) - S(A_L) - S(A_R) \right]
\label{pee2}.
\end{equation}
Note that equation~\eqref{pee2} is not directly an expression of conditional mutual information, although the two can be related, as we will explain below.

\begin{figure}[H]
    \centering
    \includegraphics[scale=0.55]{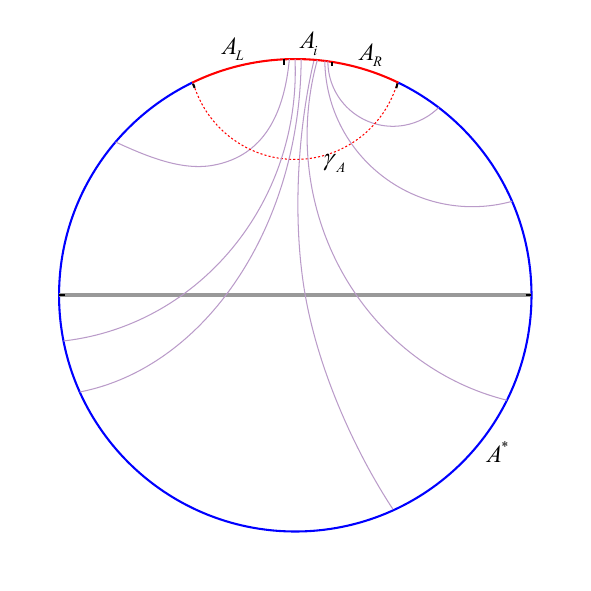}
    \caption{ The partial entanglement entropy decomposition of the planar BTZ black hole (Poincaré disk perspective). For a boundary region $A$ (marked in red), its complement with respect to the full boundary of the two-sided BTZ black hole is $ A^*$ (marked in blue). Accounting for the contributions of entanglement threads from the entire $ A^*$ region yields the partial entanglement entropy formula~\eqref{pee2} presented in previous literature.}
    \label{fig:431}
\end{figure}

We now show that the entanglement thread configuration from the two-sided BTZ perspective in fact gives a finer structure for the PEE in the single-sided BTZ scenario. As shown in FIG.~\ref{fig:431}, this is based on a very simple reason: in a global pure state, the von Neumann entropies of two complementary subsystems are equal (which does not hold for a global mixed state). Therefore, for ~\eqref{pee2}, if we denote the complement of $ A $ on the single-sided BTZ boundary $ U $ by $ \tilde{A} = U \setminus A $, and define $ A^* = \tilde{A} \cup D $ as the complement of $ A $ with respect to the full boundary of the two-sided BTZ black hole (or equivalently, the entire boundary of the Poincaré disk), then we have:
\begin{equation}
\begin{aligned}
S(A_i \cup A_R) &= S(A_L \cup A^*) \\
S(A_R) &= S(A_i \cup A_L \cup A^*)
\end{aligned}.
\end{equation}
Substituting into ~\eqref{pee2}, we obtain
\begin{equation}
s_A(A_i) = \frac{1}{2} \left[ S(A_i \cup A_L) + S(A_L \cup A^*) - S(A_L) - S(A_i \cup A_L \cup A^*) \right] 
\equiv \frac{1}{2} I(A_i, A^* \mid A_L)
\label{pee3}.
\end{equation}
From this, we see that from the global two-sided BTZ perspective, the contribution of $ A_i $ to $ S(A) $ is actually given by the total flow of entanglement threads connecting $ A^* $ to $ A_i $. The key point is to recognize that $ A^* $ includes not only $ \tilde{A} $ but also the other side $ D $. Hence, our analysis is consistent with previous studies of holographic partial entanglement entropy in the single-sided BTZ black hole. However, within our framework, from ~\eqref{pee3}, we can further write:
\begin{equation}\label{further}
s_A(A_i) = \sum_{j \in U \setminus A} F_{ij} + \sum_{\bar{j} \in D} F_{i\bar{j}}.
\end{equation}
The explicit form of the PEE expression in the planar BTZ black hole case has been computed in~\cite{Wen:2018whg,Kudler-Flam:2019oru}. Obviously, it can be verified that, according to~\eqref{further}, substituting expressions \eqref{fij2} and \eqref{nice} into it will yield exactly the same results as those in previous literature.

\section{Conclusion and Discussion}\label{sec5}

\begin{figure}
    \centering
    \begin{subfigure}[b]{0.33\textwidth}
         \centering
        \includegraphics[width=\textwidth]{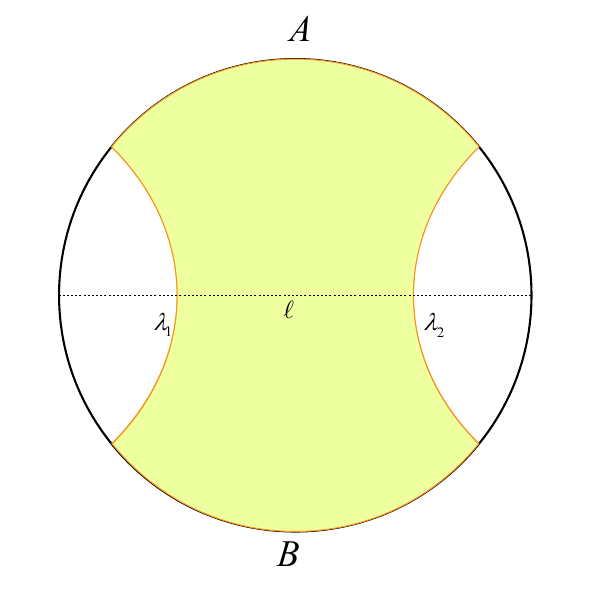}
         \caption{}
    \label{plus1}
    \end{subfigure}
    \begin{subfigure}[b]{0.27\textwidth}
         \centering
         \includegraphics[width=\textwidth]{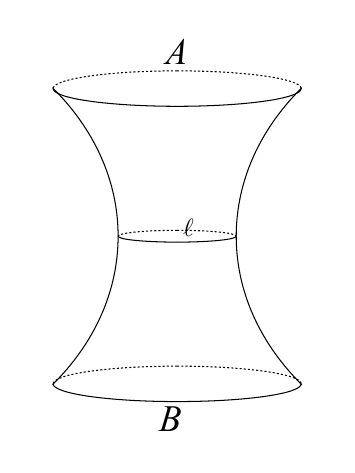}
         \caption{}
    \label{plus2}
\end{subfigure}
\begin{subfigure}[b]{0.35\textwidth}
         \centering
         \includegraphics[width=\textwidth]{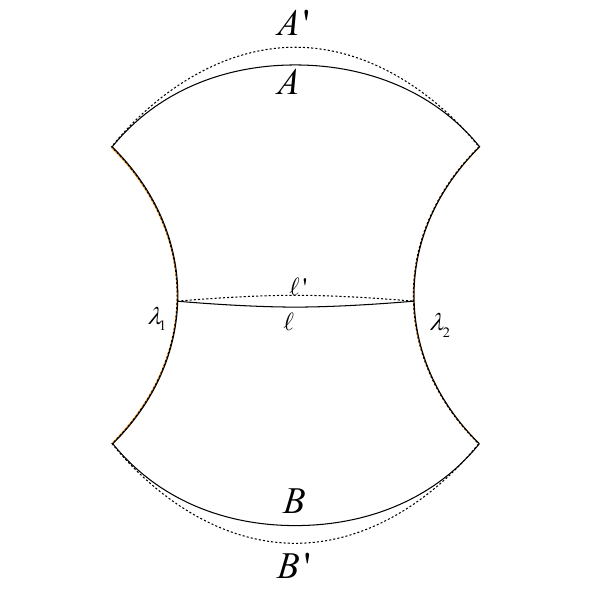}
         \caption{}
    \label{plus3}
\end{subfigure}
    \caption{Panels (a) (b) illustrate how the wormhole geometry is constructed by the quotient identification of global AdS$_3$. Panels (a) (c) illustrate how the canonical purification of a general holographic mixed state $AB$ is dual to the gluing of two copies of entanglement wedges along the minimal surfaces. Interestingly, the two perspectives may possibly be related. We leave this exploration for future work.}
    \label{plus}
\end{figure}

In this paper, we explored the entanglement structure of the BTZ black hole from an intriguing perspective — that of entanglement threads. Although there have been prior studies in this direction, such as those based on bit threads or the kinematic space approach, we revisited the problem in the framework of entanglement threads, which led to subtle differences in results compared to earlier works. More importantly, this perspective offers a new interpretation of the entanglement structure of black holes, stemming from the quantum state properties associated with entanglement threads. In summary, we have pointed out that for a finite-temperature quantum state of an infinite-size CFT dual to a planar BTZ black hole, the contribution to the holographic entanglement entropy of a subregion originates not only from the complementary region within the single-sided system, but also from the other asymptotic boundary system connected through the wormhole. Within the framework of entanglement threads, we provided precise expressions for these two contributions. By incorporating the quantum state interpretation of entanglement threads, we further analyzed the apparent RT surface phase transition in the holographic BTZ black hole setup. In fact, this phase transition signifies a structure in which entanglement threads traversing the wormhole and internal entanglement within the single-sided system together form a perfect tensor pattern with quantum error-correcting properties. Our results offer a refinement of previous analyses of the partial entanglement entropy in planar BTZ black holes.


When the spacetime dimension of the bulk is maintained at three, our analysis on the entanglement thread structure of the planar BTZ black hole can, in principle, be generalized to the case of more general bulk spacetimes. Since a more concrete scheme requires resolving some explicit technical subtleties, we leave the detailed implementation for future work. Nevertheless, at least in principle, it is possible to articulate these ideas. This involves two key points. The first key point is that every classical solution in the AdS$_3$ gravity (for example, various multi-boundary wormholes) could be constructed by the quotient identification of global pure AdS$_3$~\cite{Brill:1995jv, Aminneborg:1997pz, Brill:1998pr, Krasnov:2000zq, Skenderis:2009ju, Balasubramanian:2014hda}. As shown in Figs.~\ref{plus1} and \ref{plus2}, focusing on the static slice, for instance, the construction of the two-sided spherical BTZ geometry could be understood as the identification of the geodesics pairwise (such as the symmetric $\lambda_{1}$ and $\lambda_{2}$ in the figure) in the Poincaré disk~\cite{Banados:1992gq, Brill:1998pr}. The second key point is to maintain and further develop the idea of employing tensor network states or quantum circuit representations to characterize those slices beyond the pure AdS$_3$ case. Combining these two points, one can, according to such a recipe, use the tensor network corresponding to the original Poincaré disk to construct the holographic tensor network corresponding to other, more general slices: cut out the specified piece of the tensor network corresponding to the fundamental domain (the yellow region in the figure bounded by $\lambda_{1}$ and $\lambda_{2}$) from the original full-disk tensor network, and then glue it along those identified boundaries, i.e., contract the dangling bonds along the gluing interface. Such an idea is already familiar in previous tensor network literature, see e.g.~\cite{Czech:2015xna, Bhattacharyya:2016hbx, Peach:2017npp, Marolf:2015vma}. Then, by employing the method developed in~\cite{Lin:2025yko}, we can abstract the concept of entanglement threads within the tensor network and determine the concrete trajectories and flow fluxes of entanglement threads in the new construction. Roughly speaking, the entanglement thread flows in the original full disk are truncated at the boundary of the fundamental domain, and then will be re-glued together under the gluing operation according to some recipe, forming the new entanglement thread flows. We leave the detailed study for future work. The above line of thought can be extended to the cases of more general multi-boundary wormholes~\cite{Brill:1995jv, Aminneborg:1997pz, Brill:1998pr, Krasnov:2000zq, Skenderis:2009ju, Balasubramanian:2014hda}, since we can similarly perform such an analysis by selecting the corresponding fundamental domain on the Poincaré disk and gluing along the identified surfaces.

Another interesting and direct generalization is to extend the analysis of entanglement threads for the particular finite-temperature mixed state in this paper to the analysis of more general holographic mixed states. Here we find that the ideas in~\cite{Dutta:2019gen} concerning holographic reflected entropy and its bulk dual may be extremely useful. Considering a more general bipartite holographic quantum system in an overall mixed state (such as $AB$ in Figs.~\ref{plus1} and \ref{plus3}), \cite{Dutta:2019gen} proposed that its canonical purification can be interpreted as dual to a CPT doubling of the original bulk geometry glued to the original geometry along minimal surfaces that separate the boundary regions. Moreover, the resulting spacetime was argued to have a continuous metric, a well-defined causal structure, and to solve Einstein’s equations with a stress-energy tensor that satisfies the Null Energy Condition. Therefore, by employing reasoning similar to that in the analysis of multi-boundary wormholes, we can cut out the specified piece of the tensor network corresponding to the bipartite mixed state, and subsequently glue it to its CPT copy along the minimal surfaces to obtain the tensor network of a new bulk spacetime. We can then analyze the entanglement thread configurations therein, which should also be formed by gluing together the local entanglement thread configurations of the two glued networks. We expect that the entanglement thread configurations in the new bulk can reasonably explain holographic reflected entropy, and be applied to analyze the intrinsic correlation structure of more general holographic mixed states. It is also an interesting direction to attempt to apply entanglement thread analysis to another concept similar to reflected entropy, namely, entanglement of purification~\cite{Takayanagi:2017knl, Nguyen:2017yqw}.

Finally, let us briefly discuss the generalization to higher dimensions. The extension of our analysis of entanglement threads for the planar BTZ black hole itself to the case of dimensions higher than three is, in principle, straightforward, because many relevant tools have already been well developed, such as the previous study of bit threads in higher dimensional spaces~\cite{Freedman:2016zud, Agon:2018lwq} and higher dimensional kinematic space~\cite{Czech:2016xec, Penna:2018xqq}. Notice that our entanglement threads can be described in the language of kinematic space, where each entanglement thread can be understood as a point in kinematic space. This approach does not rely on specific dimensions, so the present analysis in principle can be directly generalized to the higher-dimensional case. However, the study of more general wormhole spacetimes and more general holographic mixed states becomes more nontrivial when extended to higher dimensions, because beyond three dimensions, generic solutions of Einstein’s equations cannot always be obtained by quotient identification of pure AdS. We leave the study of entanglement threads in various nontrivial holographic spacetimes in more general dimensions for future work.

\newpage
\begin{appendix}

\section{Appendix}

\begin{figure}
\centering
\includegraphics[width=1\linewidth]{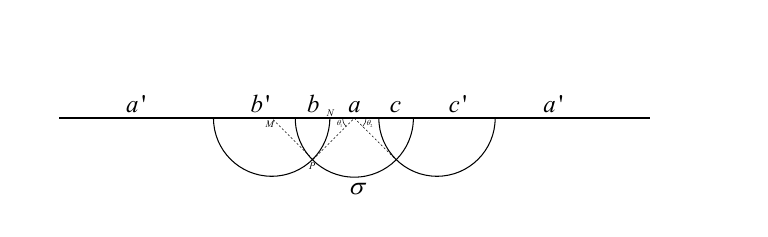}\caption{}
\label{fapp}
\end{figure}

In this appendix, we discuss a purely geometrical problem on the Poincaré disk.  For convenience, we employ the upper half-plane representation.  It is well known that, due to the existence of isometries, both the problem and the solution discussed here apply equally well to the Poincaré disk representation as well as to the two-sided BTZ black hole representation.  

According to the letter markings in the FIG.~\ref{fapp}, we formulate the problem as follows:  Consider the geodesic $\Sigma $ extended from the endpoints of $U=b\cup a\cup c$ (corresponding to the upper boundary $U$ in the main text), which serves as the horizon surface $\Sigma$ in the main text.  From the two endpoints of the subregion $a$ (corresponding to $A_i$ in the main text), draw two geodesics outward such that they intersect $\Sigma$ orthogonally (In fact, these two geodesics are precisely the elementary reference lines in the main text.).  The two intersection points then determine a segment $\sigma$ on $\Sigma$ (corresponding to $\sigma_i$ in the main text).  We aim to prove (where $d$ denotes the hyperbolic length)  
\begin{equation}
d_\sigma = \frac{1}{2} I(a:D \mid b), \label{ques}
\end{equation}
where, as shown in the figure,  
$D = b' \cup a' \cup c'$ is the complement of $U$ (corresponding to the lower boundary $D$ in the main text), and  $I(a:D \mid b) \equiv d_{ab} + d_{ac} - d_b - d_c.
$\footnote{Note that we have slightly modified the system of units so that the entropy values match exactly with the geodesic lengths.}  

The proof is as follows. As shown in the figure, let the radius of $U$ be $R$.  Due to the crucial orthogonality condition, it is evident that $MN=MP$ implies  
\begin{equation}
R \tan \theta_{1} = \frac{R}{\cos \theta_{1}} - R + b.
\end{equation}
Excluding the trivial solution $\theta=0$, another real solution is obtained as  
\begin{align}
\cos \theta_{1} 
&= -\frac{2(k-1)}{k^{2}-2k+2} \\
&= \frac{-2}{(k-1)+\tfrac{1}{(k-1)}} 
   \in \left[ -\tfrac{1}{\sqrt{2}},\,0 \right) \label{th1},
\end{align}
where $k=b/R$. Similarly, we obtain the solution for $\theta_2$:  
\begin{equation}
\cos \theta_{2} = -\frac{2(m-1)}{m^{2}-2m+2} \label{th2}.
\end{equation}
Having derived explicit expressions for $\theta_1$ and $\theta_2$,  we can integrate the line element $ds$ from $\theta_1$ to $\pi-\theta_2$ to compute the length of $\sigma$:  
\begin{equation}
d_\sigma = \int_{\theta_{1}}^{\pi - \theta_{2}} \frac{dx^{2} + dy^{2}}{y^{2}}.
\end{equation}
By parameterizing $x=R\cos\theta$, $y=R\sin\theta$, and defining $\theta'_2=\pi-\theta_2$, we obtain  
\begin{align}
d_\sigma
&= \int_{\theta_{1}}^{\theta_{2}'} \frac{1}{\sin \theta}\, d\theta \\
&= \tfrac{1}{2} \log \!\left( \frac{1-\cos\theta}{1+\cos\theta} \right)\Bigg|_{\theta_{1}}^{\theta_{2}'}.
\end{align}
Substituting \eqref{th1} and \eqref{th2}, we immediately get  
\begin{align}
d_\sigma 
&= \tfrac{1}{2} \log \!\left( 
   \frac{(m^{2}-2m+2) - 2(m-1)}{(m^{2}-2m+2) + 2(m-1)} 
   \cdot 
   \frac{(k^{2}-2k+2) - 2(k-1)}{(k^{2}-2k+2) + 2(k-1)} 
   \right) \\
&= \log \!\left( \frac{2R-c}{c} \cdot \frac{2R-b}{b} \right) \label{result}.
\end{align}
From the well-known geodesic formula (with curvature radius set to 1 and UV cutoff $\varepsilon$):  $d_l = 2\log \frac{l}{\varepsilon }$, it is evident that the result \eqref{result} matches exactly the right-hand side of \eqref{ques}.  
Thus, we have completed the proof.

\end{appendix}

\newpage{\pagestyle{empty}\cleardoublepage}


\begin{thebibliography}{100}

	\bibitem{Swingle:2009bg} B.~Swingle,
	``Entanglement Renormalization and Holography,''
	Phys. Rev. D \textbf{86}, 065007 (2012)
	[arXiv:0905.1317 [cond-mat.str-el]].
	
	\bibitem{Swingle:2012wq} B.~Swingle,
	``Constructing holographic spacetimes using entanglement renormalization,''
	[arXiv:1209.3304 [hep-th]].

\bibitem{Pastawski:2015qua}
F.~Pastawski, B.~Yoshida, D.~Harlow and J.~Preskill,
``Holographic quantum error-correcting codes: Toy models for the bulk/boundary correspondence,''
JHEP \textbf{06}, 149 (2015)
[arXiv:1503.06237 [hep-th]].

\bibitem{Vidal:2007hda} G.~Vidal,
	``Entanglement Renormalization,''
	Phys. Rev. Lett. \textbf{99}, no.22, 220405 (2007)
	[arXiv:cond-mat/0512165 [cond-mat]].
\bibitem{Vidal:2008zz} G.~Vidal,
	``Class of Quantum Many-Body States That Can Be Efficiently Simulated,''
	Phys. Rev. Lett. \textbf{101}, 110501 (2008)
	[arXiv:quant-ph/0610099 [quant-ph]].
\bibitem{Vidal:2015} E.~Glen, G.~Vidal,
 ``Tensor network renormalization yields the multiscale entanglement renormalization ansatz,''
 Phys. Rev. Lett. \textbf{115},200401 (2015).
[arXiv:cond-mat/1502.05385 [cond-mat]].



	\bibitem{Maldacena:1997re} J.~M.~Maldacena,
``The Large N limit of superconformal field theories and supergravity,''
Adv. Theor. Math. Phys. \textbf{2}, 231-252 (1998)
[arXiv:hep-th/9711200 [hep-th]].


	\bibitem{Gubser:1998bc} S.~S.~Gubser, I.~R.~Klebanov and A.~M.~Polyakov,
``Gauge theory correlators from noncritical string theory,''
Phys. Lett. B \textbf{428}, 105-114 (1998)
[arXiv:hep-th/9802109 [hep-th]].


	\bibitem{Witten:1998qj} E.~Witten,
``Anti-de Sitter space and holography,''
Adv. Theor. Math. Phys. \textbf{2}, 253-291 (1998)
[arXiv:hep-th/9802150 [hep-th]].



	\bibitem{Freedman:2016zud} M.~Freedman and M.~Headrick,
	``Bit threads and holographic entanglement,''
	Commun. Math. Phys. \textbf{352}, no.1, 407-438 (2017)
	[arXiv:1604.00354 [hep-th]].
	\bibitem{Cui:2018dyq} S.~X.~Cui, P.~Hayden, T.~He, M.~Headrick, B.~Stoica and M.~Walter,
	``Bit Threads and Holographic Monogamy,''
	Commun. Math. Phys. \textbf{376}, no.1, 609-648 (2019)
	[arXiv:1808.05234 [hep-th]].
	\bibitem{Headrick:2017ucz} M.~Headrick and V.~E.~Hubeny,
	``Riemannian and Lorentzian flow-cut theorems,''
	Class. Quant. Grav. \textbf{35}, no.10, 10 (2018)
	[arXiv:1710.09516 [hep-th]].
	\bibitem{Headrick:2022nbe} M.~Headrick and V.~E.~Hubeny,
``Covariant bit threads,''
JHEP \textbf{07}, 180 (2023)
[arXiv:2208.10507 [hep-th]].

\bibitem{Headrick:2020gyq}
M.~Headrick, J.~Held and J.~Herman,
``Crossing Versus Locking: Bit Threads and Continuum Multiflows,''
Commun. Math. Phys. \textbf{396}, no.1, 265-313 (2022)
[arXiv:2008.03197 [hep-th]].

\bibitem{Agon:2018lwq}
C.~A.~Ag\'on, J.~De Boer and J.~F.~Pedraza,
``Geometric Aspects of Holographic Bit Threads,''
JHEP \textbf{05}, 075 (2019)
[arXiv:1811.08879 [hep-th]].

\bibitem{Kudler-Flam:2019oru}
J.~Kudler-Flam, I.~MacCormack and S.~Ryu,
``Holographic entanglement contour, bit threads, and the entanglement tsunami,''
J. Phys. A \textbf{52}, no.32, 325401 (2019)
[arXiv:1902.04654 [hep-th]].

\bibitem{Lin:2025yko}
Y.~Y.~Lin,
``The thread embodiment of holographic quantum entanglement,''
[arXiv:2501.10691 [hep-th]].

\bibitem{Lin:2023jah}
Y.~Y.~Lin and J.~Zhang,
``Holographic coarse-grained states and the necessity of perfect entanglement,''
Phys. Rev. D \textbf{109}, no.12, 126012 (2024)
[arXiv:2312.14498 [hep-th]].
\bibitem{Lin:2022flo}
Y.~Y.~Lin and J.~C.~Jin,
``Thread/State correspondence: from bit threads to qubit threads,''
JHEP \textbf{02}, 245 (2023)
[arXiv:2210.08783 [hep-th]].
\bibitem{Lin:2022agc}
Y.~Y.~Lin and J.~C.~Jin,
``Thread/State correspondence: the qubit threads model of holographic gravity,''
[arXiv:2208.08963 [hep-th]].
\bibitem{Lin:2020yzf}
Y.~Y.~Lin, J.~R.~Sun and Y.~Sun,
``Bit thread, entanglement distillation, and entanglement of purification,''
Phys. Rev. D \textbf{103}, no.12, 126002 (2021)
[arXiv:2012.05737 [hep-th]].
\bibitem{Lin:2023orb}
Y.~Y.~Lin,
``Distilled density matrices of holographic partial entanglement entropy from thread-state correspondence,''
Phys. Rev. D \textbf{108}, no.10, 106010 (2023)
[arXiv:2305.02895 [hep-th]].
\bibitem{Lin:2021hqs}
Y.~Y.~Lin, J.~R.~Sun and J.~Zhang,
``Deriving the PEE proposal from the locking bit thread configuration,''
JHEP \textbf{10}, 164 (2021)
[arXiv:2105.09176 [hep-th]].


\bibitem{Jahn:2019nmz}
A.~Jahn, M.~Gluza, F.~Pastawski and J.~Eisert,
``Majorana dimers and holographic quantum error-correcting codes,''
Phys. Rev. Research. \textbf{1}, 033079 (2019)
[arXiv:1905.03268 [hep-th]].
\bibitem{Yan:2019vzp}
H.~Yan,
``Geodesic string condensation from symmetric tensor gauge theory: a unifying framework of holographic toy models,''
Phys. Rev. B \textbf{102}, no.16, 161119 (2020)
[arXiv:1911.01007 [cond-mat.str-el]].
\bibitem{Lin:2024dho}
J.~Lin, Y.~Lu and Q.~Wen,
``Partial entanglement network and bulk geometry reconstruction in AdS/CFT,''
[arXiv:2401.07471 [hep-th]].
\bibitem{Lin:2023rxc}
J.~Lin, Y.~Lu and Q.~Wen,
``Geometrizing the partial entanglement entropy: from PEE threads to bit threads,''
JHEP \textbf{2024}, no.02, 191 (2024)
[arXiv:2311.02301 [hep-th]].

\bibitem{Lin:2023hzs}
Y.~Y.~Lin, J.~Zhang and J.~C.~Jin,
``Entanglement islands read perfect-tensor entanglement,''
JHEP \textbf{04}, 113 (2024)
[arXiv:2312.14486 [hep-th]].
\bibitem{Lin:2022aqf}
Y.~Y.~Lin, J.~R.~Sun, Y.~Sun and J.~C.~Jin,
``The PEE aspects of entanglement islands from bit threads,''
JHEP \textbf{07}, 009 (2022)
[arXiv:2203.03111 [hep-th]].


\bibitem{Wen:2024uwr}
Q.~Wen, M.~Xu and H.~Zhong,
``Partial entanglement entropy threads in the island phase,''
Phys. Rev. D \textbf{111}, no.4, 046027 (2025)
[arXiv:2408.13535 [hep-th]].

\bibitem{Caggioli:2024uza}
S.~Caggioli, F.~Gentile, D.~Seminara and E.~Tonni,
``Holographic thermal entropy from geodesic bit threads,''
JHEP \textbf{07}, 088 (2024)
[arXiv:2403.03930 [hep-th]].

\bibitem{Mintchev:2022fcp}
M.~Mintchev and E.~Tonni,
``Modular conjugations in 2D conformal field theory and holographic bit threads,''
JHEP \textbf{12}, 149 (2022)
[arXiv:2209.03242 [hep-th]].











	\bibitem{Ryu:2006bv} S.~Ryu and T.~Takayanagi,
``Holographic derivation of entanglement entropy from AdS/CFT,''
Phys. Rev. Lett. \textbf{96}, 181602 (2006)
[arXiv:hep-th/0603001 [hep-th]].


	\bibitem{Ryu:2006ef} S.~Ryu and T.~Takayanagi,
``Aspects of Holographic Entanglement Entropy,''
JHEP \textbf{08}, 045 (2006)
[arXiv:hep-th/0605073 [hep-th]].


	\bibitem{Hubeny:2007xt} V.~E.~Hubeny, M.~Rangamani and T.~Takayanagi,
``A Covariant holographic entanglement entropy proposal,''
JHEP \textbf{07}, 062 (2007)
[arXiv:0705.0016 [hep-th]].

  
    
  
\bibitem{Czech:2015qta} B.~Czech, L.~Lamprou, S.~McCandlish and J.~Sully,
``Integral Geometry and Holography,''
JHEP \textbf{10}, 175 (2015)
[arXiv:1505.05515 [hep-th]].
\bibitem{Czech:2015kbp} B.~Czech, L.~Lamprou, S.~McCandlish and J.~Sully,
    ``Tensor Networks from Kinematic Space,''
    JHEP \textbf{07}, 100 (2016)
    [arXiv:1512.01548 [hep-th]].



\bibitem{Bao:2015bfa}
N.~Bao, S.~Nezami, H.~Ooguri, B.~Stoica, J.~Sully and M.~Walter,
``The Holographic Entropy Cone,''
JHEP \textbf{09}, 130 (2015)
[arXiv:1505.07839 [hep-th]].

\bibitem{Hubeny:2018ijt}
V.~E.~Hubeny, M.~Rangamani and M.~Rota,
``The holographic entropy arrangement,''
Fortsch. Phys. \textbf{67}, no.4, 1900011 (2019)
[arXiv:1812.08133 [hep-th]].

\bibitem{Hubeny:2018trv}
V.~E.~Hubeny, M.~Rangamani and M.~Rota,
``Holographic entropy relations,''
Fortsch. Phys. \textbf{66}, no.11-12, 1800067 (2018)
[arXiv:1808.07871 [hep-th]].

\bibitem{HernandezCuenca:2019wgh}
S.~Hern\'andez Cuenca,
``Holographic entropy cone for five regions,''
Phys. Rev. D \textbf{100}, no.2, 026004 (2019)
[arXiv:1903.09148 [hep-th]].

\bibitem{Vidal:2014aal}
G.~Vidal and Y.~Chen,
``Entanglement contour,''
J. Stat. Mech. \textbf{2014}, no.10, P10011 (2014)
[arXiv:1406.1471 [cond-mat.str-el]].

\bibitem{Wen:2019iyq}
Q.~Wen,
``Formulas for Partial Entanglement Entropy,''
Phys. Rev. Res. \textbf{2}, no.2, 023170 (2020)
[arXiv:1910.10978 [hep-th]].


\bibitem{Wen:2018whg}
Q.~Wen,
``Fine structure in holographic entanglement and entanglement contour,''
Phys. Rev. D \textbf{98}, no.10, 106004 (2018)
[arXiv:1803.05552 [hep-th]].






\bibitem{Asplund:2016koz}
C.~T.~Asplund, N.~Callebaut and C.~Zukowski,
``Equivalence of Emergent de Sitter Spaces from Conformal Field Theory,''
JHEP \textbf{09}, 154 (2016)
[arXiv:1604.02687 [hep-th]].

\bibitem{Levay:2020rzs}
P.~L{\'e}vay and B.~Boldis,
``Scanning spacetime with patterns of entanglement,''
Phys. Rev. D \textbf{101}, no.6, 066021 (2020)
[arXiv:2001.07923 [hep-th]].

\bibitem{Boldis:2021snw}
B.~Boldis and P.~L{\'e}vay,
``Cluster algebraic description of entanglement patterns for the BTZ black hole,''
Phys. Rev. D \textbf{105}, no.4, 046020 (2022)
[arXiv:2108.10638 [hep-th]].


\bibitem{Brown:1986nw}
J.~D.~Brown and M.~Henneaux,
``Central Charges in the Canonical Realization of Asymptotic Symmetries: An Example from Three-Dimensional Gravity,''
Commun. Math. Phys. \textbf{104}, 207-226 (1986)

\bibitem{Calabrese:2004eu}
P.~Calabrese and J.~L.~Cardy,
``Entanglement entropy and quantum field theory,''
J. Stat. Mech. \textbf{0406}, P06002 (2004)
[arXiv:hep-th/0405152 [hep-th]].

\bibitem{Maldacena:2001kr}
J.~M.~Maldacena,
``Eternal black holes in anti-de Sitter,''
JHEP \textbf{04}, 021 (2003)
[arXiv:hep-th/0106112 [hep-th]].

\bibitem{Israel:1976ur}
W.~Israel,
``Thermo field dynamics of black holes,''
Phys. Lett. A \textbf{57}, 107-110 (1976)


\bibitem{Banados:1992wn}
M.~Banados, C.~Teitelboim and J.~Zanelli,
``The Black hole in three-dimensional space-time,''
Phys. Rev. Lett. \textbf{69}, 1849-1851 (1992)
[arXiv:hep-th/9204099 [hep-th]].

\bibitem{Banados:1992gq}
M.~Banados, M.~Henneaux, C.~Teitelboim and J.~Zanelli,
``Geometry of the (2+1) black hole,''
Phys. Rev. D \textbf{48}, 1506-1525 (1993)
[erratum: Phys. Rev. D \textbf{88}, 069902 (2013)]
[arXiv:gr-qc/9302012 [gr-qc]].




\bibitem{Molina-Vilaplana:2011ydi}
J.~Molina-Vilaplana and P.~Sodano,
``Holographic View on Quantum Correlations and Mutual Information between Disjoint Blocks of a Quantum Critical System,''
JHEP \textbf{10}, 011 (2011)
[arXiv:1108.1277 [quant-ph]].

\bibitem{Matsueda:2012xm}
H.~Matsueda, M.~Ishihara and Y.~Hashizume,
``Tensor network and a black hole,''
Phys. Rev. D \textbf{87}, no.6, 066002 (2013)
[arXiv:1208.0206 [hep-th]].

\bibitem{Molina-Vilaplana:2014mna}
J.~Molina-Vilaplana and J.~Prior,
``Entanglement, Tensor Networks and Black Hole Horizons,''
Gen. Rel. Grav. \textbf{46}, no.11, 1823 (2014)
[arXiv:1403.5395 [hep-th]].

\bibitem{Bao:2015uaa}
N.~Bao, C.~Cao, S.~M.~Carroll, A.~Chatwin-Davies, N.~Hunter-Jones, J.~Pollack and G.~N.~Remmen,
``Consistency conditions for an AdS multiscale entanglement renormalization ansatz correspondence,''
Phys. Rev. D \textbf{91}, no.12, 125036 (2015)
[arXiv:1504.06632 [hep-th]].



\bibitem{Facchi:2008ora}
P.~Facchi, G.~Florio, G.~Parisi and S.~Pascazio,
``Maximally multipartite entangled states,''
Phys. Rev. A \textbf{77}, no.6, 060304 (2008)
\bibitem{Helwig:2012nha}
W.~Helwig, W.~Cui, A.~Riera, J.~I.~Latorre and H.~K.~Lo,
``Absolute Maximal Entanglement and Quantum Secret Sharing,''
Phys. Rev. A \textbf{86}, 052335 (2012)
[arXiv:1204.2289 [quant-ph]].
\bibitem{Helwig:2013ckb}
W.~Helwig and W.~Cui,
``Absolutely Maximally Entangled States: Existence and Applications,''
[arXiv:1306.2536 [quant-ph]].
\bibitem{Helwig:2013qoq}
W.~Helwig,
``Absolutely Maximally Entangled Qudit Graph States,''
[arXiv:1306.2879 [quant-ph]].


\bibitem{Almheiri:2014lwa}
A.~Almheiri, X.~Dong and D.~Harlow,
``Bulk Locality and Quantum Error Correction in AdS/CFT,''
JHEP \textbf{04}, 163 (2015)
[arXiv:1411.7041 [hep-th]].
\bibitem{Dong:2016eik}
X.~Dong, D.~Harlow and A.~C.~Wall,
``Reconstruction of Bulk Operators within the Entanglement Wedge in Gauge-Gravity Duality,''
Phys. Rev. Lett. \textbf{117}, no.2, 021601 (2016)
[arXiv:1601.05416 [hep-th]].


\bibitem{Brill:1995jv}
D.~R.~Brill,
``Multi - black hole geometries in (2+1)-dimensional gravity,''
Phys. Rev. D \textbf{53}, 4133-4176 (1996)
[arXiv:gr-qc/9511022 [gr-qc]].

\bibitem{Aminneborg:1997pz}
S.~Aminneborg, I.~Bengtsson, D.~Brill, S.~Holst and P.~Peldan,
``Black holes and wormholes in (2+1)-dimensions,''
Class. Quant. Grav. \textbf{15}, 627-644 (1998)
[arXiv:gr-qc/9707036 [gr-qc]].

\bibitem{Brill:1998pr}
D.~Brill,
``Black holes and wormholes in (2+1)-dimensions,''
Lect. Notes Phys. \textbf{537}, 143 (2000)
[arXiv:gr-qc/9904083 [gr-qc]].

\bibitem{Krasnov:2000zq}
K.~Krasnov,
``Holography and Riemann surfaces,''
Adv. Theor. Math. Phys. \textbf{4}, 929-979 (2000)
[arXiv:hep-th/0005106 [hep-th]].

\bibitem{Skenderis:2009ju}
K.~Skenderis and B.~C.~van Rees,
``Holography and wormholes in 2+1 dimensions,''
Commun. Math. Phys. \textbf{301}, 583-626 (2011)
[arXiv:0912.2090 [hep-th]].

\bibitem{Balasubramanian:2014hda}
V.~Balasubramanian, P.~Hayden, A.~Maloney, D.~Marolf and S.~F.~Ross,
``Multiboundary Wormholes and Holographic Entanglement,''
Class. Quant. Grav. \textbf{31}, 185015 (2014)
[arXiv:1406.2663 [hep-th]].



\bibitem{Czech:2015xna}
B.~Czech, G.~Evenbly, L.~Lamprou, S.~McCandlish, X.~L.~Qi, J.~Sully and G.~Vidal,
``Tensor network quotient takes the vacuum to the thermal state,''
Phys. Rev. B \textbf{94}, no.8, 085101 (2016)
[arXiv:1510.07637 [cond-mat.str-el]].

\bibitem{Bhattacharyya:2016hbx}
A.~Bhattacharyya, Z.~S.~Gao, L.~Y.~Hung and S.~N.~Liu,
``Exploring the Tensor Networks/AdS Correspondence,''
JHEP \textbf{08}, 086 (2016)
[arXiv:1606.00621 [hep-th]].

\bibitem{Peach:2017npp}
A.~Peach and S.~F.~Ross,
``Tensor Network Models of Multiboundary Wormholes,''
Class. Quant. Grav. \textbf{34}, no.10, 105011 (2017)
[arXiv:1702.05984 [hep-th]].

\bibitem{Marolf:2015vma}
D.~Marolf, H.~Maxfield, A.~Peach and S.~F.~Ross,
``Hot multiboundary wormholes from bipartite entanglement,''
Class. Quant. Grav. \textbf{32}, no.21, 215006 (2015)
[arXiv:1506.04128 [hep-th]].

\bibitem{Dutta:2019gen}
S.~Dutta and T.~Faulkner,
``A canonical purification for the entanglement wedge cross-section,''
JHEP \textbf{03}, 178 (2021)
[arXiv:1905.00577 [hep-th]].

\bibitem{Takayanagi:2017knl}
T.~Takayanagi and K.~Umemoto,
``Entanglement of purification through holographic duality,''
Nature Phys. \textbf{14}, no.6, 573-577 (2018)
[arXiv:1708.09393 [hep-th]].

\bibitem{Nguyen:2017yqw}
P.~Nguyen, T.~Devakul, M.~G.~Halbasch, M.~P.~Zaletel and B.~Swingle,
``Entanglement of purification: from spin chains to holography,''
JHEP \textbf{01}, 098 (2018)
[arXiv:1709.07424 [hep-th]].

\bibitem{Czech:2016xec}
B.~Czech, L.~Lamprou, S.~McCandlish, B.~Mosk and J.~Sully,
``A Stereoscopic Look into the Bulk,''
JHEP \textbf{07}, 129 (2016)
[arXiv:1604.03110 [hep-th]].

\bibitem{Penna:2018xqq}
R.~F.~Penna and C.~Zukowski,
``Kinematic space and the orbit method,''
JHEP \textbf{07}, 045 (2019)
[arXiv:1812.02176 [hep-th]].





\end{thebibliography}
\end{document}